\documentclass[12pt]{article}
\usepackage{amsmath,bm,amsfonts,color,amsthm, amssymb}
\usepackage[right=1.0in, left=1.0in, top=1.2in, bottom=2.0in]{geometry}
\usepackage{graphicx}
\usepackage{natbib}
\usepackage{xcolor}
\usepackage{enumitem}
\usepackage{subcaption,siunitx,booktabs}
\usepackage[citecolor=blue,urlcolor=blue,allcolors=blue]{hyperref}
\usepackage{arydshln}

\usepackage{titlesec}
\titlelabel{\thetitle.\quad}

\usepackage{tikz}
\usepackage{transparent}

\usetikzlibrary{positioning}
\usetikzlibrary{shadows}
\usepgflibrary{shapes}

\usepackage{mathtools}

\usepackage{apptools}
\AtAppendix{\counterwithin{lemma}{section}}

\AtAppendix{\counterwithin{proposition}{section}}

\usepackage{titlesec}

\titleformat*{\section}{\large\bfseries}
\titleformat*{\subsection}{\large\bfseries}
\titleformat*{\subsubsection}{\large\bfseries}
\titleformat*{\paragraph}{\large\bfseries}
\titleformat*{\subparagraph}{\large\bfseries}

\usepackage{titlesec}
\titlelabel{\thetitle\quad}

\theoremstyle{definition}

\usepackage{amsmath}
\usepackage{amssymb}
\usepackage{amsfonts}
\usepackage{multirow}
\usepackage{amsthm}
\usepackage[ruled,vlined]{algorithm2e}
\usepackage{setspace}

\newtheorem{theorem}{Theorem}

\theoremstyle{definition}

\newtheorem{remark}{Remark}

\numberwithin{equation}{section}

\title{An Unbiased Predictor for Skewed Response Variable with Measurement Error in Covariate}

  \author{Sepideh Mosaferi\thanks{
    Contact the corresponding author at \url{smosaferi@umass.edu}}, Malay Ghosh, and Shonosuke Sugasawa \hspace{.2cm}\\
    {\it University of Massachusetts Amherst, University of Florida} \\
    {\it and Keio University}}

\begin{document}

\date{}

\maketitle
\begin{abstract}

We introduce a new small area predictor when the Fay-Herriot normal error model is fitted to a logarithmically transformed response variable, and the covariate is measured with error. This framework has been previously studied by \cite{mosaferi2020+}. The empirical predictor given in their manuscript cannot perform uniformly better than the direct estimator. Our proposed predictor in this manuscript is unbiased and can perform uniformly better than the one proposed in \cite{mosaferi2020+}. We derive an approximation of the mean squared error (MSE) for the predictor. The prediction intervals based on the MSE suffer from coverage problems. Thus, we propose a non-parametric bootstrap prediction interval which is more  accurate. 
This problem is of great interest in small area applications since statistical agencies and agricultural surveys are often asked to produce estimates of right skewed variables with covariates measured with errors. With Monte Carlo simulation studies and two Census Bureau's data sets, we demonstrate the superiority of our proposed methodology.

\vspace{0.25cm}
\noindent {\it Key words and phrases:}
Bayes estimator, prediction interval, transformation. 

\end{abstract}

\section{Introduction} \label{sec:Introduction}

Small area estimation concerns producing estimates or predictions of means, totals or quantiles for each of a finite collection of geographic regions, where there are a small number of sampled units in each individual region (area).     
Classical models used in small area estimation take the form of mixed linear models that result from the concatenation of a model for error in direct sample-based estimators for each area and an additional model that connects areas through the use of covariates and area-specific random effects.  

These {\it linking models} take the direct estimators to be linear combinations of covariates and random effects.  We focus here on what is called the {\it area level model} (\cite{ghosh1994small}, \cite{pfeffermann2013new}, and \cite{rao2015small}, Chap. 4) which uses covariates at the level of the areas. Recently, \cite{mosaferi2020+} proposed a model of the Fay-Herriot type and developed an empirical predictor for small area quantities that they are right skewed. 

A complication that arises is that the area-level covariates to be used can be the result of survey sampling (see for instance \cite{ybarra2008small}), thus producing a small area model with measurement error in the covariates. The predictor given in their manuscript cannot perform uniformly better than the direct estimator because of bias issues. In this manuscript, we propose a new unbiased predictor, which can perform uniformly better than that of \cite{mosaferi2020+} and the direct estimator. 

Much work with  small area models has been devoted to the estimation of MSE. 
\cite{prasad1990estimation} derived a closed form approximation for the estimator of the MSE of an empirical Bayes (EB) predictor under the assumption of normality. 
\cite{jiang2002unified} proposed a jackknife estimator based on a decomposition of MSE where the leading term is approximately unbiased and does not depend on the area-specific random effects.  \cite{butar2003measures} developed a bootstrap estimator of MSE.

Here, we derive an approximation of the MSE for the predictor as well as the jackknife estimator of MSE.
Prediction intervals based on these suffer from inadequate coverage probabilities. In order to address this shortcoming, we develop prediction intervals based on a non-parametric bootstrap method. In the rest of this section, we list some of the previous works in the literature and highlight our contributions.

\subsection{Prior Work} \label{sec.prior_work}
\cite{molina2018empirical} and \cite{berg2014small} worked on the log-transformation model and proposed an EB predictor for the value of the variable of interest for out-of-sample individuals (and for small area means) in a  nested-error regression model where no measurement error is assumed present in the covariates.
The analytical MSE given in \cite{molina2018empirical} has a complex form. Thus, the authors proposed a parametric bootstrap procedure for estimation of the uncertainty following \cite{butar2003measures}. 
\cite{slud2006mean} proposed a large sample approximation to the MSE of the predictor for the transformed Fay-Herriot model without measurement error. 

\subsection{Our Contributions} \label{sec.contribution}

In this paper, we make several contributions to the literature. First, unlike the earlier works given in Section \ref{sec.prior_work}, we assume the available covariate in the model is measured with error. Second, we propose a new small area predictor for the skewed response variable in the original scale at the area-level instead of unit-level under presence of measurement error in covariate and make comparisons with the earlier predictor proposed by \cite{mosaferi2020+}. 
Third, we explain how to estimate the unknown parameters using unbiased score functions from the marginal likelihood.  Finally, we derive an approximation for the MSE of our proposed predictor and develop prediction intervals based on nonparametric bootstrap techniques. 

The rest of the paper is organized as follows. 
In Section \ref{sec:FHmodel}, we apply the \cite{fay1979estimates} model to the transformed data with measurement error in covariate and formulate the problem. In Section \ref{sec:Predictors}, we derive a new predictor for the response variable in the proposed modeling framework. 
In Section \ref{sec:unknown_pars}, we explain how to estimate the unknown parameters in the model. In Section \ref{sec:MSE}, we derive an estimator of the MSE of the predictor.

In Section \ref{sec:CIs}, we construct non-parametric bootstrap prediction intervals. In Section \ref{sec:Simulations}, using a Monte Carlo simulation study, we make comparisons with other predictors given in the literature. 
In Section \ref{sec:Realdata}, we illustrate our methodology using two data sets from the Census Bureau. 
The related discussions and possible extensions are given in Section \ref{sec:discussion}. 
Technical details and additional numerical results are in the Supplementary Material. All the R code implementing the proposed methodology is available at Github repository \url{https://github.com/SepidehMosaferi/UnbiasedPredictor_SkewedData}. 

\section{Transformed Fay-Herriot Model with Measurement Errors} \label{sec:FHmodel}

Assume response variables $y_i$ ($i=1,...,m$) are right skewed. Thus, the log transformation of $y_i$ can stabilize the variation. 
Let $z_i=\log(y_i)$ and $z_i=\phi_i+e_i$, where $\phi_i=\log(\theta_i)$ such that $\theta_i$ is unknown and $e_i \overset{\text{ind}}{\sim} N(0, \psi_i)$ is the sampling error.
We further define $\phi_i=\beta_0+\beta_1 x_i+ v_i$, where the covariate $x_i$ is not observed, and what one observes is the $W_i$. The linking error is $v_i \overset{\text{iid}}{\sim} N(0, \sigma^2_v)$.
The error terms ($e_i,v_i$) are mutually independent per each $i$-th small area. 

Then, the transformed Fay-Herriot model with measurement error in covariate can be presented with the following hierarchical set-up
\begin{align} \label{eq:FH_hierarchy}
z_i | \phi_i & \overset{\text{ind}}{\sim} N(\phi_i, \psi_i) \nonumber\\
\phi_i & \overset{\text{ind}}{\sim} N(\beta_0+\beta_1 x_i, \sigma^2_v) \nonumber\\
W_i & \overset{\text{ind}}{\sim} N(x_i,C_i), \quad i=1,...,m.
\end{align}

The $(z_i,\phi_i)$ are assumed to be independent of the $W_i$. This is because the former constitutes the sampling and linking model, while the later brings in the measurement error part. 
Here, following \cite{ybarra2008small}, $\sigma^2_v$ is unknown but the sampling variances $\psi_i$ and $C_i$ are assumed to be known, which can be obtained from the asymptotic variances of transformed direct estimates (see, \cite{carter1974empirical}, \cite{efron1975data}, and \cite{fay1979estimates}). 

The primary parameter of interest in the original scale is $\theta_i \equiv \exp(\beta_0 + \beta_1 x_i+v_i)$.  
Prediction of $\phi_i=\log (\theta_i)$, where $\log (\theta_i) \equiv \beta_0+\beta_1 x_i+v_i$, is identical to the problem of \cite{ybarra2008small}. 

We will adopt an empirical Bayes approach for estimation of the $\theta_i$. We will assume a flat prior for all the $x_i$'s, but then estimate $\beta_0$, $\beta_1$, and $\sigma^2_v$ from the resulting marginal likelihood. To this end, first observe that writing $\gamma^\star_i=\sigma^2_v/(\sigma^2_v+\psi_i)$, the conditional posterior distributions
\begin{equation*}
\phi_i |\beta_0, \beta_1, \sigma^2_v, x_i, z_i, W_i  \overset{\text{ind}}{\sim} N(\gamma^\star_i z_i+(1-\gamma^\star_i)(\beta_0+\beta_1 x_i), \gamma^\star_i \psi_i),    
\end{equation*}
and $x_i|\beta_0, \beta_1, \sigma^2_v, z_i, W_i$ are mutually independent, where
\begin{equation*}
\pi(x_i|\beta_0, \beta_1, \sigma^2_v, z_i, W_i) \propto (\sigma^2_v+\psi_i)^{-1/2} \exp \Big[-\frac{1}{2} \Big\{\frac{(z_i -\beta_0 - \beta_1 x_i)^2}{\sigma^2_v+\psi_i} + \frac{(W_i-x_i)^2}{C_i} \Big\} \Big].   
\end{equation*}

Next we use the identity, 
\begin{align} \label{eq.identity}
& (z_i-\beta_0-\beta_1 x_i)^2/(\sigma^2_v+\psi_i)+(W_i-x_i)^2/C_i  = \{\beta_1^2 C_i+\sigma^2_v+\psi_i\}^{-1} \times (z_i - \beta_0 - \beta_1 W_i)^2 \nonumber\\
& \quad + \{\beta_1^2/(\sigma^2_v+\psi_i)+C_i^{-1}\} 
 \times \Big(x_i - \Big\{\frac{\beta_1 (z_i - \beta_0)}{\sigma^2_v+\psi_i}+\frac{W_i}{C_i}\Big\} \Big/ \Big\{\frac{\beta_1^2}{\sigma^2_v+\psi_i}+\frac{1}{C_i}\Big\} \Big)^2.
\end{align}
Now writing $S_i(\beta_1,\sigma^2_v)=\beta_1^2 C_i+\sigma^2_v+\psi_i$, one gets
\begin{equation*}
x_i | \beta_0, \beta_1, \sigma^2_v, z_i, W_i \overset{\text{ind}}{\sim} N \Big[\frac{\beta_1 C_i(z_i-\beta_0)+(\sigma^2_v+\psi_i)W_i}{S_i(\beta_1,\sigma^2_v)}, \frac{\sigma^2_v+\psi_i}{S_i(\beta_1,\sigma^2_v)}\Big].     
\end{equation*}

Accordingly, 
\begin{align*}
E(\phi_i | \beta_0, \beta_1, \sigma^2_v, z_i, W_i) & = E\Big[E(\phi_i | \beta_0, \beta_1, \sigma^2_v, x_i, z_i, W_i) | \beta_0, \beta_1, \sigma^2_v, z_i, W_i \Big] \\
& = \frac{\sigma^2_v}{\sigma^2_v+\psi_i} z_i + \frac{\psi_i}{\sigma^2_v+\psi_i} \Big[\beta_0+\beta_1 \Big\{\frac{\beta_1 C_i(z_i-\beta_0)+(\sigma^2_v+\psi_i) W_i}{S_i(\beta_1,\sigma^2_v)} \Big\} \Big] 
\end{align*}

\begin{align*}
& = \frac{\sigma^2_v \, S_i(\beta_1,\sigma^2_v) + \psi_i \beta_1^2 C_i}{(\sigma^2_v+\psi_i)S_i(\beta_1,\sigma^2_v)} z_i+\frac{\psi_i}{\sigma^2_v+\psi_i} \Big[\frac{\beta_0 (\sigma^2_v+\psi_i)}{S_i(\beta_1,\sigma^2_v)}+\frac{\beta_1(\sigma^2_v+\psi_i) W_i}{S_i(\beta_1,\sigma^2_v)} \Big] \\
& = \frac{\beta_1^2 C_i+\sigma^2_v}{S_i(\beta_1,\sigma^2_v)} z_i + \frac{\psi_i(\beta_0+\beta_1 W_i)}{S_i(\beta_1,\sigma^2_v)} = \tilde{\gamma}_i z_i + (1-\tilde{\gamma}_i) (\beta_0+\beta_1 W_i),
\end{align*}
where $\tilde{\gamma}_i \equiv (\beta_1^2 C_i+\sigma^2_v)/S_i(\beta_1,\sigma^2_v)$. Further,
\begin{align*}
V(\phi_i| \beta_0, \beta_1, \sigma^2_v, z_i, W_i) & = E \Big[V(\phi_i| \beta_0, \beta_1, \sigma^2_v, x_i, z_i, W_i) | \beta_0, \beta_1, \sigma^2_v, z_i, W_i \Big] \\
& \quad + V \Big[E(\phi_i| \beta_0, \beta_1, \sigma^2_v, x_i, z_i, W_i) | \beta_0, \beta_1, \sigma^2_v, z_i, W_i \Big] \\
& = \gamma_i^\star \psi_i+(1-\gamma_i^\star)^2 \beta_1^2 \frac{(\sigma^2_v+\psi_i)C_i}{S_i(\beta_1,\sigma^2_v)} \\
& = \frac{\sigma^2_v \psi_i}{\sigma^2_v+\psi_i} + \frac{\psi_i^2}{(\sigma^2_v+\psi_i)^2} \beta_1^2 C_i \frac{\sigma^2_v+\psi_i}{S_i(\beta_1,\sigma^2_v)} \\
& = \frac{1}{\sigma^2_v+\psi_i} \Big[\sigma^2_v \psi_i+\frac{\psi_i^2 \beta_1^2 C_i}{S_i(\beta_1,\sigma^2_v)} \Big] \\
& = \frac{\psi_i}{(\sigma^2_v+\psi_i)S_i(\beta_1,\sigma^2_v)} \Big[\sigma^2_v(\beta_1^2 C_i+\sigma^2_v+\psi_i) + \beta_1^2 C_i \psi_i \Big] \\
& = \frac{\psi_i}{(\sigma^2_v+\psi_i)S_i(\beta_1,\sigma^2_v)} \Big[\sigma^2_v (\sigma^2_v+\psi_i) + \beta_1^2 C_i (\sigma^2_v+\psi_i) \Big] = \tilde{\gamma}_i \psi_i.
\end{align*}

This leads to the Bayes estimator 
\begin{align*} \label{eq:PredA}
\tilde{\theta}_{i,A} & = E[\exp(\phi_i) | \beta_0, \beta_1, \sigma^2_v, z_i, W_i] \nonumber \\
& = \exp (\tilde{\gamma}_i z_i + (1-\tilde{\gamma}_i) (\beta_0+\beta_1 W_i) + \tilde{\gamma}_i \psi_i/2),
\end{align*}
proposed by \cite{mosaferi2020+}.
Also,

\begin{align*}
V[\exp(\phi_i) | \beta_0, \beta_1, \sigma^2_v, z_i, W_i] & = \exp\{\tilde{\gamma}_i \psi_i  \} [\exp\{\tilde{\gamma}_i \psi_i  \}-1] \\
& \quad \times \exp\{2[\tilde{\gamma}_i z_i + (1-\tilde{\gamma}_i) (\beta_0+\beta_1 W_i)] \}.
\end{align*}

Further, from (\ref{eq.identity}), the marginal likelihood of $(\beta_0,\beta_1,\sigma^2_v)$ is given by
\begin{equation} \label{eq:marginal}
L_M(\beta_0,\beta_1,\sigma^2_v) = \prod_{i=1}^{m} S_i^{-1/2}(\beta_1,\sigma^2_v) \, \exp \Big[-\frac{1}{2} \sum_{i=1}^{m} \frac{\tau_i^2(\beta_0,\beta_1)}{S_i(\beta_1,\sigma^2_v)} \Big],
\end{equation} 
where we define $\tau_i(\beta_0,\beta_1)=z_i-\beta_0-\beta_1 W_i$. 
In the reminder of this paper, we will work with the score functions based on the marginal likelihood (\ref{eq:marginal}) to estimate the vector of unknown parameters.

\section{Optimal Predictor in the Original Scale} \label{sec:Predictors} 

It is clear that $E(\tilde{\theta}_{i,A})=E(\theta_i)=\exp(\beta_0+\beta_1 x_i + \frac{1}{2} \sigma^2_v)$ when $C_i=0$. The above equality, however, is not true in general. To this end, we prove the following theorem.

\begin{theorem} \label{theorem.bias}
$E(\tilde{\theta}_{i,A})=\exp(\beta_0+\beta_1 x_i+\frac{1}{2} 
\{ \tilde{\gamma}_i (\sigma^2_v+\psi_i)+(1-\tilde{\gamma}_i) \beta_1^2 C_i \})$.

\vspace{0.15cm}
\noindent \textit{Proof:} See Supplementary Material.
\end{theorem} 

In view of Theorem \ref{theorem.bias}, 
\begin{align*}
\frac{E(\tilde{\theta}_{i,A})}{E(\theta_i)} & = \exp \Big(\frac{1}{2} \{\tilde{\gamma}_i (\sigma^2_v+\psi_i) + (1-\tilde{\gamma}_i) \beta_1^2 C_i - \sigma^2_v \} \Big) \\
& = \exp \Big( \frac{1}{2} d_i \Big), \quad \text{(say)}
\end{align*}
that is
$$
E \Big\{ \tilde{\theta}_{i,A} \exp \Big(- \frac{1}{2} d_i \Big) \Big\} = E(\theta_i).
$$
Therefore, our proposed unbiased predictor is
\begin{equation} \label{eq:AandB}
\tilde{\theta}_{i,B} = \tilde{\theta}_{i,A} \exp \Big(- \frac{1}{2} \, d_i \Big).
\end{equation}  
With some algebra, $d_i$ can be simplified as $d_i= 2 \psi_i \beta_1^2 C_i / S_i(\beta_1,\sigma^2_v)$.

\begin{remark}
When $C_i=0$, the true value of $x_i$ can be used for the optimal predictor. 
By substituting $C_i=0$ into $\tilde{\theta}_{i,B}$, the optimal predictor is $\tilde{\theta}_{i,B}^{0}= \exp(\gamma_i^\star z_i + (1-\gamma_i^\star) (\beta_0+\beta_1x_i)+\gamma_i^\star \psi_i/2)$, which is same as the predictor given in \cite{slud2006mean} and is also identical to predictor $\tilde{\theta}_{i,A}$. 
\end{remark}

\section{Estimating the Unknown Parameters} \label{sec:unknown_pars}

We are interested in obtaining estimates of the vector of unknown parameters $\boldsymbol{\omega}=(\beta_0,\beta_1,\sigma^2_v)'$. Note that we do not directly use the partial derivatives of the marginal likelihood given in expression (\ref{eq:marginal}) since they are not unbiased.
We denote the log-likelihood of $L_M(.)$ in (\ref{eq:marginal}) by 
$\ell_M(\boldsymbol{\omega})$. 
The score functions of $\ell_M(\boldsymbol{\omega})$ can be defined as follows: 
\begin{equation*}
U(\boldsymbol{\omega})= (U_1(\boldsymbol{\omega}), U_2(\boldsymbol{\omega}), U_3(\boldsymbol{\omega}))' = \Big(\frac{\partial \ell_M(\boldsymbol{\omega})}{\partial \beta_0},  \frac{\partial \ell_M(\boldsymbol{\omega})}{\partial \beta_1}, \frac{\partial \ell_M(\boldsymbol{\omega})}{\partial \sigma^2_v}  \Big)'.
\end{equation*} 

These score functions are biased for estimating the unknown parameters $\boldsymbol{\omega}$. Thus, we define the unbiased score functions as follows

\begin{equation} \label{eq:unbiasdscore}
\tilde{U}(\boldsymbol{\omega})= (\tilde{U}_1(\boldsymbol{\omega}), \tilde{U}_2(\boldsymbol{\omega}), \tilde{U}_3(\boldsymbol{\omega}))' = U(\boldsymbol{\omega}) - E[U(\boldsymbol{\omega})],
\end{equation} 
such that $E[\tilde{U}(\boldsymbol{\omega})]=0$.
Original score functions for $U(\boldsymbol{\omega})$ are
\begin{align*}
U_1(\boldsymbol{\omega})& = \sum_{i=1}^{m} S_i^{-1}(\beta_1,\sigma^2_v) \tau_i(\beta_0,\beta_1), \\
U_2(\boldsymbol{\omega}) & = - \sum_{i=1}^{m} S_i^{-1}(\beta_1,\sigma^2_v) \beta_1 C_i  + \sum_{i=1}^{m} S_i^{-1}(\beta_1,\sigma^2_v) W_i \tau_i(\beta_0,\beta_1) \\
& \quad + \sum_{i=1}^{m} S_i^{-2}(\beta_1,\sigma^2_v) \tau_i^2(\beta_0,\beta_1)\beta_1 C_i, \\
U_3(\boldsymbol{\omega}) & = - \frac{1}{2} \sum_{i=1}^{m} S_i^{-1}(\beta_1,\sigma^2_v) + \frac{1}{2} \sum_{i=1}^{m} S_i^{-2}(\beta_1,\sigma^2_v) \tau_i^2(\beta_0,\beta_1),
\end{align*}
and their expected values are
\begin{align*}
E[U_1(\boldsymbol{\omega})] = 0, \quad E[U_2(\boldsymbol{\omega})] = - \sum_{i=1}^{m} S_i^{-1}(\beta_1,\sigma^2_v) \beta_1 C_i, \quad E[U_3(\boldsymbol{\omega})]=0,
\end{align*}
where we emphasize that $E[W_i \tau_i(\beta_0,\beta_1)]= - \beta_1 C_i$.

Using expression (\ref{eq:unbiasdscore}), the unbiased score functions for estimating $\boldsymbol{\omega}$ are
\begin{align} \label{eq:finalunbiasdscore}
\tilde{U}_1(\boldsymbol{\omega})& = \sum_{i=1}^{m} S_i^{-1}(\beta_1,\sigma^2_v) \tau_i(\beta_0,\beta_1)=0,  \nonumber \\
\tilde{U}_2(\boldsymbol{\omega}) & = \sum_{i=1}^{m} S_i^{-1}(\beta_1,\sigma^2_v) W_i \tau_i(\beta_0,\beta_1) + \sum_{i=1}^{m} S^{-2}_i(\beta_1,\sigma^2_v) \tau_i^2(\beta_0,\beta_1) \beta_1 C_i = 0,  \nonumber \\
\tilde{U}_3(\boldsymbol{\omega}) & = - \frac{1}{2} \sum_{i=1}^{m} S_i^{-1}(\beta_1,\sigma^2_v) + \frac{1}{2} \sum_{i=1}^{m} S^{-2}_i(\beta_1,\sigma^2_v) \tau_i^2(\beta_0,\beta_1)=0.
\end{align}

One can solve equations given in (\ref{eq:finalunbiasdscore}) numerically to find the estimates of unknown parameters.
Given the current estimate $\boldsymbol{\omega}^{(r)}$ of $\boldsymbol{\omega}$, by replacing $S_i(\beta_1,\sigma_v^2)$ with $S_i(\beta_1^{(r)},\sigma_v^{2(r)})$ and $\beta_1$ with $\beta_1^{(r)}$ in $ \tilde{U}_1(\boldsymbol{\omega})$, the solution of $\beta_0$ is given by 
$$
\beta_0=\left\{\sum_{i=1}^mS_i^{-1}(\beta_1^{(r)},\sigma_v^{2(r)}) \right\}^{-1}\left\{\sum_{i=1}^mS_i^{-1}(\beta_1^{(r)},\sigma_v^{2(r)})(z_i-\beta_1^{(r)}W_i)\right\}.
$$

Similarly, by replacing $S_i(\beta_1,\sigma_v^2)$ with $S_i(\beta_1^{(r)},\sigma_v^{2(r)})$ and $\beta_0$ with $\beta_0^{(r)}$ in $ \tilde{U}_2(\boldsymbol{\omega})$, the solution of $\beta_1$ is given by 
\begin{align*}
\beta_1 & =\left\{\sum_{i=1}^m S_i^{-1}(\beta_1^{(r)},\sigma_v^{2(r)})W_i^2-\sum_{i=1}^m S_i^{-2}(\beta_1^{(r)},\sigma_v^{2(r)})\tau_i^2(\beta_0^{(r)},\beta_1^{(r)})C_i \right\}^{-1} \\
& \quad \times \left\{\sum_{i=1}^mS_i^{-1}(\beta_1^{(r)},\sigma_v^{2(r)})W_i(z_i-\beta_0^{(r)})\right\}.
\end{align*}
Therefore, we can develop the iterative Algorithm \ref{algo1} to solve the estimating equations.

\begin{algorithm}[]
\SetAlgoLined
1. Set the initial value $\boldsymbol{\omega}^{(0)}$ and $r=0$. \\
\For{$r=1, \ldots,R$}{
2. Update $\beta_0$ as 
\[ \beta_0^{(r+1)}=\Big\{\sum_{i=1}^mS_i^{-1}(\beta_1^{(r)},\sigma_v^{2(r)}) \Big\}^{-1}\Big\{\sum_{i=1}^mS_i^{-1}(\beta_1^{(r)},\sigma_v^{2(r)})(z_i-\beta_1^{(r)}W_i)\Big\}. \]

3. Update $\beta_1$ as 
\begin{align*}
\beta_1^{(r+1)}
&=\Big\{\sum_{i=1}^mS_i^{-1}(\beta_1^{(r)},\sigma_v^{2(r)})W_i^2- \sum_{i=1}^m S_i^{-2}(\beta_1^{(r)},\sigma_v^{2(r)})\tau_i^2(\beta_0^{(r+1)},\beta_1^{(r)})C_i \Big\}^{-1} \\
&  \ \ \ \ \ \ \ 
\times \Big\{\sum_{i=1}^mS_i^{-1}(\beta_1^{(r)},\sigma_v^{2(r)})W_i(z_i-\beta_0^{(r+1)})\Big\}.
\end{align*}

4. Update $\sigma_v^2$ (obtain $\sigma_v^{2(r+1)}$) by solving the equation 
\begin{align*}
-\frac{1}{2} \sum_{i=1}^m S_i^{-1}(\beta_1^{(r+1)},\sigma_v^2)+\frac{1}{2} \sum_{i=1}^mS_i^{-2}(\beta_1^{(r+1)}, \sigma_v^2 )\tau_i^2(\beta_0^{(r+1)}, \beta_1^{(r+1)})=0.
\end{align*}

5. If $\|\boldsymbol{\omega}^{(r+1)}-\boldsymbol{\omega}^{(r)}\|<\varepsilon$ with tolerance $\varepsilon>0$, $\boldsymbol{\omega}^{(r+1)}$ is the final estimate; otherwise, set $r=r+1$ and go back to Step 2. 

}
\KwResult{$\hat{\boldsymbol{\omega}}=(\hat{\beta}_0,\hat{\beta}_1,\hat{\sigma}^2_v)'$.}
\caption{Iterative algorithm for solving the unbiased estimating equations  (\ref{eq:finalunbiasdscore})}
\label{algo1}
\end{algorithm}

\begin{remark}
Note that the updating process $\beta_1$ in Algorithm \ref{algo1} is similar to the modified least squares estimator used in \cite{ybarra2008small} in which a fixed weight is used instead of $S_i^{-1}(\beta_1,\sigma_v^2)$.
Thus, the updating process in the iterative algorithm can be regarded as iteratively reweighted least squares. 
\end{remark}

\clearpage
\begin{theorem} \label{theorem.I_matrix}
Define $ \tilde{\sigma}^2_{ci}:=C_i(\sigma^2_v+\psi_i)S_i^{-1}(\beta_1,\sigma^2_v)$. Based on the properties of the unbiased estimating functions, one can obtain $[\hat{\boldsymbol{\omega}}-\boldsymbol{\omega}] \xrightarrow{D} N_3 (\boldsymbol{0}, I_{\boldsymbol{\omega}}^{-1})$ as $m \rightarrow \infty$, where
\begin{equation*}
I_{\boldsymbol{\omega}} = \begin{bmatrix}
\sum_{i=1}^{m}S_i^{-1}(\beta_1,\sigma^2_v) & \sum_{i=1}^{m} S_i^{-1}(\beta_1,\sigma^2_v) x_i & 0 \\
\sum_{i=1}^{m} S_i^{-1}(\beta_1,\sigma^2_v) x_i  & \sum_{i=1}^{m} S_i^{-1}(\beta_1,\sigma^2_v) (x_i^2+ \tilde{\sigma}^2_{ci} )& 0\\
0 & 0 & \frac{1}{2} \sum_{i=1}^{m} S_i^{-2}(\beta_1,\sigma^2_v)
\end{bmatrix}.
\end{equation*}

\vspace{0.15cm}
\noindent \textit{Proof:} See Supplementary Material.
\end{theorem}

\section{Mean Squared Error Formulae} \label{sec:MSE}
In this Section, we find an expression for the MSE of $\tilde{\theta}_{i,B}$ which is correct up to $O(m^{-1/2})$ as well as the jackknife estimator of MSE for $\tilde{\theta}_{i,B}^E:=\tilde{\theta}_{i,B}(\hat{\boldsymbol{\omega}})$. The MSE of the empirical predictor B is
\begin{align} \label{MSE}
\text{MSE}(\tilde{\theta}_{i,B}^E) = E \Big[(\tilde{\theta}_{i,B}-\theta_i)^2 \Big] + E \Big[(\tilde{\theta}_{i,B}^E-\tilde{\theta}_{i,B})^2 \Big]:= R_{1i}+R_{2i},
\end{align} 
where $R_{1i}$ is equal to
\begin{align*}
R_{1i} & = E \Big[(\tilde{\theta}_{i,B}-\theta_i)^2 \Big]=E \Big[(\theta_i - \tilde{\theta}_{i,A})^2 \Big] + E \Big[(\tilde{\theta}_{i,A}- \tilde{\theta}_{i,B})^2 \Big]. 
\end{align*} 
Firstly,
\begin{align*}
E \Big[(\theta_i - \tilde{\theta}_{i,A})^2 \Big]  & = E \Big[ E\Big\{(\theta_i - \tilde{\theta}_{i,A})^2 | \beta_0, \beta_1 , \sigma^2_v, z_i, W_i \Big\}\Big] = E \Big[V(\theta_i) | \beta_0, \beta_1, \sigma^2_v, z_i, W_i \Big] \\
& = E \Big[E(\theta_i^2 | \beta_0, \beta_1, \sigma^2_v, z_i, W_i) - \tilde{\theta}_{i,A}^2 \Big] = E(\theta_i^2) - E(\tilde{\theta}_{i,A}^2).
\end{align*}

\noindent Secondly, $E \Big[ (\tilde{\theta}_{i,A} - \tilde{\theta}_{i,B})^2 \Big] = E \Big[ \tilde{\theta}_{i,A}^2 \{ 1- \exp(-\frac{1}{2} \, d_i) \}^2 \Big].$
By combining the expressions, we have
\begin{equation*}
E \Big[ (\tilde{\theta}_{i,B} - \theta_i)^2 \Big] = E(\theta_i^2) + E(\tilde{\theta}_{i,A}^2) \Big\{ -2 \exp(- \frac{1}{2} \, d_i) + \exp(-d_i) \Big\}. 
\end{equation*}

Additionally,
\begin{align*}
E(\tilde{\theta}_{i,A}^2) & = E\Big[\exp \Big\{2 \tilde{\gamma}_i z_i+2(1-\tilde{\gamma}_i)(\beta_0+\beta_1 W_i)+\tilde{\gamma}_i \psi_i \Big\}\Big] \\
& = \exp \Big[2 \Big\{\tilde{\gamma}_i (\beta_0+\beta_1 x_i)+(1-\tilde{\gamma}_i) (\beta_0+\beta_1 x_i) \Big\} \Big] \\
& \quad \times \exp \Big[2 \tilde{\gamma}_i^2 (\sigma^2_v+\psi_i) + 2 (1-\tilde{\gamma}_i)^2 \beta_1^2 C_i + \tilde{\gamma}_i \psi_i \Big] \\
& = \exp \Big[2(\beta_0+\beta_1 x_i+\sigma^2_v) \Big] \exp(-2 \sigma^2_v) \exp(- \tilde{\gamma}_i \psi_i) \\
& \quad \times \exp \Big[2 \Big\{\tilde{\gamma}_i^2 (\sigma^2_v+\psi_i) + (1-\tilde{\gamma}_i)^2 \beta_1^2 C_i + \tilde{\gamma}_i \psi_i \Big\} \Big]. 
\end{align*}
Note $\tilde{\gamma}_i^2 (\sigma^2_v+\psi_i) + (1-\tilde{\gamma}_i)^2 \beta_1^2 C_i + \tilde{\gamma}_i \psi_i=d_i+r^2$. Thus,
\begin{equation*}
E(\tilde{\theta}_{i,A}^2) = E(\theta_i^2) \exp \Big\{ 2 d_i - \tilde{\gamma}_i \psi_i \Big\}.
\end{equation*}

As a result
\begin{align} \label{R1i}
R_{1i} = E \Big[ (\tilde{\theta}_{i,B} - \theta_i)^2 \Big] & = E(\theta_i^2) + E(\theta_i^2) \exp \Big\{ 2 d_i - \tilde{\gamma}_i \psi_i \Big\} \Big\{ -2 \exp \Big(- \frac{1}{2} \, d_i \Big) + \exp(-d_i) \Big\} \nonumber\\
& = E(\theta_i^2) \Big\{1 -2 \exp \Big(\frac{3}{2} \, d_i - \tilde{\gamma}_i \psi_i \Big) + \exp[d_i- \tilde{\gamma}_i \psi_i] \Big\} \nonumber \\
& := M_{1i}(\boldsymbol{\omega}) M_{2i}(\boldsymbol{\omega}),
\end{align} 
where $E(\theta_i^2)= \exp[2 (\beta_0+\beta_1 x_i + \sigma^2_v)]$.
We define $M_{1i}(\boldsymbol{\omega})$ as follows

\begin{align*}
M_{1i}(\boldsymbol{\omega}) := E(\theta_i^2) = E[ \exp(2 \phi_i)] = \exp[2\beta_0+2\beta_1 x_i+ 2 \sigma^2_v]. 
\end{align*}
Now note that $E[\exp(2 z_i)] = \exp[2 \beta_0+2\beta_1 x_i+ 2 \sigma^2_v+2 \psi_i]$. In order to find an unbiased estimator for $M_{1i}(\boldsymbol{\omega})$, one can define 
$\hat{M}_{1i}(\hat{\boldsymbol{\omega}}) := \exp[2(z_i-\psi_i)],$
so that $E[\hat{M}_{1i}(\hat{\boldsymbol{\omega}})]= M_{1i}(\boldsymbol{\omega})$.

Now, we have
$E[\hat{M}_{1i}(\hat{\boldsymbol{\omega}})-M_{1i}(\boldsymbol{\omega}) ]^2 = E[\hat{M}^2_{1i}(\hat{\boldsymbol{\omega}})] - M^2_{1i}(\boldsymbol{\omega}),$
so that
\begin{align*}
E[\hat{M}^2_{1i}(\hat{\boldsymbol{\omega}})] & = E[\exp \{4 (z_i-\psi_i) \}] = \exp(- 4 \psi_i) E[\exp(4 z_i)] \\
& = \exp(- 4 \psi_i) \exp[4 \beta_0+4 \beta_1 x_i+8 \sigma^2_v+8 \psi_i].
\end{align*}
Therefore, 
\begin{align*}
E[\hat{M}_{1i}(\hat{\boldsymbol{\omega}})-M_{1i}(\boldsymbol{\omega}) ]^2 & = \exp(- 4 \psi_i) \exp[4 \beta_0+ 4 \beta_1 x_i + 8 \sigma^2_v + 8 \psi_i] \\
& \quad - \exp(- 4 \psi_i) \exp[4 \beta_0+4 \beta_1 x_i + 4 \sigma^2_v+4 \psi_i] \\
& = \exp(- 4 \psi_i) \exp[4 \beta_0+4 \beta_1 x_i+8 \sigma^2_v + 8 \psi_i] [1- \exp(- 4 \sigma^2_v- 4 \psi_i)].
\end{align*}
One can estimate $E[\hat{M}_{1i}(\hat{\boldsymbol{\omega}})-M_{1i}(\boldsymbol{\omega}) ]^2$ by $\Lambda_i(\hat{\boldsymbol{\omega}})$ defined as follows:
\begin{equation*}
\Lambda_i(\hat{\boldsymbol{\omega}}) := \exp(- 4 \psi_i) \exp(4 z_i) [1- \exp(- 4 \hat{\sigma}^2_v- 4 \psi_i)].
\end{equation*}

For $M_{2i}(\boldsymbol{\omega})$ in expression (\ref{R1i}), we have
\begin{equation*}
M_{2i}(\boldsymbol{\omega}) = 1 -2 \exp \Big[\frac{3}{2} \, d_i - \tilde{\gamma}_i \psi_i \Big] + \exp[d_i- \tilde{\gamma}_i \psi_i].
\end{equation*}
Define
\begin{align*}
E[M^2_{2i}(\hat{\boldsymbol{\omega}})- M_{2i}^2(\boldsymbol{\omega})] & := E[M_{2i}(\hat{\boldsymbol{\omega}})- M_{2i}(\boldsymbol{\omega})]^2 + 2 M_{2i}(\boldsymbol{\omega}) E[M_{2i}(\hat{\boldsymbol{\omega}})- M_{2i}(\boldsymbol{\omega})],
\end{align*}
where the terms can be expanded as
\begin{align*}
\text{(i)} \quad & E[M_{2i}(\hat{\boldsymbol{\omega}})- M_{2i}(\boldsymbol{\omega})]^2 \\ & = E \Big[\Big\{ \exp(\hat{d}_i - \hat{\tilde{\gamma}}_i \psi_i) - \exp(d_i - \tilde{\gamma}_i \psi_i) \Big\} - 2 \Big\{ \exp\Big(\frac{3}{2} \hat{d}_i - \hat{\tilde{\gamma}}_i \psi_i \Big) - \exp \Big(\frac{3}{2} d_i - \gamma_i \psi_i \Big) \Big\}\Big]^2 \\
& = O(m^{-1}), \quad \text{following Theorem \ref{theorem.I_matrix}, and} \\
\text{(ii)} \quad & E|M_{2i}(\hat{\boldsymbol{\omega}})- M_{2i}(\boldsymbol{\omega})| \leq E^{1/2}[M_{2i}(\hat{\boldsymbol{\omega}}) - M_{2i}(\boldsymbol{\omega})]^2 = O(m^{-1/2}).
\end{align*}

The quantity $R_{1i}$ can be estimated with
\begin{align*}
\hat{R}_{1i} = M_{2i}^2(\hat{\boldsymbol{\omega}}) \Lambda_i(\hat{\boldsymbol{\omega}}) & = [1 -2 \exp \Big(\frac{3}{2} \hat{d}_i - \hat{\tilde{\gamma}}_i \psi_i \Big) + \exp(\hat{d}_i - \hat{\tilde{\gamma}}_i \psi_i)]^2 \\
& \quad \times \exp(- 4 \psi_i) \exp(4 z_i) [1- \exp(- 4 \hat{\sigma}^2_v- 4 \psi_i)],
\end{align*}
where it has the property that its bias and variance vanish with an order $O(m^{-1/2})$. Details of derivations are given in the Supplementary Material.

In general, there is no closed form expression available for the term $R_{2i}$ in (\ref{MSE}). One can use the jackknife technique to estimate it as well as the bias of $\hat{R}_{1i}$ for $R_{1i}$. Therefore, the jackknife estimator of $\text{MSE}(\tilde{\theta}_{i,B}^{E})$ is
\begin{equation*} \label{mse_J}
mse_J(\tilde{\theta}_{i,B}^{E}) = \hat{R}_{1i,J}+\hat{R}_{2i,J},
\end{equation*}
where $\hat{R}_{1i,J} = \hat{R}_{1i} - \frac{m-1}{m} \sum_{j=1}^{m} (\hat{R}_{1i(-j)} - \hat{R}_{1i})$ and
$\hat{R}_{2i,J} = \frac{m-1}{m} \sum_{j=1}^{m} (\tilde{\theta}_{i(-j),B}^E-\tilde{\theta}_{i,B}^E)^2$.

Under the regularity conditions 1--3 given in the Supplementary Material, one can show that $E\{\hat{R}_{1i,J} \}=R_{1i}+O(m^{-1})$ and $E \{ \hat{R}_{2i,J} \}=R_{2i}+o(m^{-1})$ as $m \rightarrow \infty$. 
Thus, $E \{mse_J(\tilde{\theta}_{i,B}^{E})\}= \text{MSE}(\tilde{\theta}_{i,B}^{E}) + O(m^{-1})$. The proof follows along the same lines of \cite{mosaferi2020+}, and hence we omit it. 

The analytical approximation for the MSE of predictor $\tilde{\theta}_{i,B}$ has a complex form. As we will find from our simulations, constructing intervals based on that as well as the jackknife estimator of MSE perform poorly in terms of coverage or length. Additionally, jackknife MSE might yield negative values. Thus, one might prefer to use resampling procedures such as bootstrap to construct the intervals as they are easier with better interpretation and coverage property.

\section{Non-parametric Bootstrap Prediction Intervals} 
\label{sec:CIs}

In this Section, we propose a non-parametric bootstrap approach to approximate the entire distribution of predictor B. We use the percentiles of the bootstrap histogram to obtain highly accurate prediction intervals in terms of coverage.
For this purpose, we repeatedly draw samples from the original observed sample.

We construct the bootstrap distribution of predictor B ($\tilde{\theta}_{i,B}$) based on the observed data $(W_i,y_i)$ for $i=1,...,m$ such that $\tilde{\theta}_{i,B}^{*}=\tilde{\theta}_{i,B}(W_i^{*},y_i^{*})$, where $(W_i^{*},y_i^{*})$ is obtained from the resampling observed pairs $(W_i,y_i)$; i.e., $\{(W_{j_1},y_{j_1}), (W_{j_2},y_{j_2}), ..., (W_{j_m},y_{j_m}) \}$ where $j_1, j_2, ..., j_m$ is a random sample drawn with replacement from $\{1, 2, ..., m \}$. Each bootstrap sample gives a non-parametric bootstrap replication of $\tilde{\theta}_{i,B}$ denoted by $\tilde{\theta}_{i,B}^*$. 
After repeating the bootstrap distribution $``BT"$ times, we obtain $\tilde{\theta}_{i,B}^{*(1)}, \tilde{\theta}_{i,B}^{*(2)}, ..., \tilde{\theta}_{i,B}^{*(BT)}$. 

We use the upper and lower $\alpha/2$ quantiles of these $BT$ numbers as the prediction interval for $\theta$. 
Specifically, we propose to use the interval
\begin{equation} \label{eq.BootCI}
\hat{I}_{i,\alpha}= [ \ell_{i,\alpha}, u_{i,\alpha}]=[\hat{G}^{-1}_i(\alpha/2),\hat{G}_i^{-1}(1-\alpha/2)],
\end{equation}
where 
$\hat{G}_i(t)= \frac{1}{BT} \sum_{bt=1}^{BT} I (\tilde{\theta}_{i,B}^{*(bt)} \leq t).$
In the above expression, $\hat{G}_i(t)$ is the cumulative distribution function (CDF) of $BT$ bootstrap replications, and we let $BT=2000$ in the application and simulation studies.

Using functional delta method (\cite{van2000asymptotic}, Chap. 20) and Theorem \ref{theorem.I_matrix}, $(\tilde{\theta}_{i,B}-\theta_{i}) \xrightarrow{D} N(0,\sigma^2)$ for a constant $\sigma^2>0$. Let's consider the sampling distribution of standardized $\tilde{\theta}_{i,B}$ (i.e. $T_i=\{\tilde{\theta}_{i,B}-\theta_i \}/\sigma_i$) defined as $G_i(t) \equiv P(T_i \leq t)$ for $t \in \mathbb{R}$. We can obtain the bootstrap estimator of $G_i(t)$ from the bootstrap version $T_i^*=\{\tilde{\theta}_{i,B}^{*}-\tilde{\theta}_{i,B} \}/\sigma_i^*$ of $T_i$ defined as $\hat{G}_i^*(t) \equiv P_{*}(T_i^* \leq t)$ for $t \in \mathbb{R}$. As $m \rightarrow \infty$ and the bootstrap replications $BT$ increase, the probability distribution $\hat{G}_i^*(t)$ weakly converges to $G_i(t)$, i.e., $\hat{G}_i^*(t) \rightarrow G_i(t)$ for all the continuity points $t$ of $G_i$ (see, \cite{hall2013bootstrap}, Chaps. 3 and 4) and under the assumption of $\sigma_i^* / \sigma _i \xrightarrow{P} 1$. 

One can use the subsequence arguments and the correspondence between convergence almost surely along subsequences and convergence in probability to show the distance between distributions $\hat{G}_i^*$ and $G_i$ goes to zero in probability. Thus, based on Polya's theorem, $\sup_{t}|\hat{G}^*_i(t)-G_i(t)|= \sup_{t} |\{\hat{G}^*_i(t)- \Phi(t)\} - \{G_i(t)- \Phi(t) \}| \xrightarrow{P} 0$ for $t \in \mathbb{R}$ and as $m \rightarrow \infty$ (\cite{lahiri2003resampling}, Chap. 2), where $\Phi(t)$ is the CDF of standard normal, and $\hat{G}^*_i(.)$ is a consistent estimator of $G_i(.)$ under the regularity conditions 1--2 given in the Supplementary Material and assuming $E[y_i^2] \, , \, E[||x_i||^2]< \infty$. 
When model (\ref{eq:FH_hierarchy}) is correctly specified,
  
\begin{align*}
P(|\sigma_i^{-1}(\theta_i-\tilde{\theta}_{i,B}^*)| \geq z_{\alpha/2}) 
& \leq P(|\sigma_i^{-1}(\theta_i-\tilde{\theta}_{i,B})| \geq z_{\alpha/2}) +
P(|\sigma_i^{*-1}(\tilde{\theta}_{i,B}-\tilde{\theta}_{i,B}^*)| \geq z_{\alpha/2}) \\
& = \alpha+o(1).
\end{align*}
This states that the prediction interval $\hat{I}_{i,\alpha}$ given in (\ref{eq.BootCI}) is asymptotically valid.

\section{Simulation Studies} 
\label{sec:Simulations}

We perform a simulation study to compare the performance of several predictors. For this purpose, we generate data from the model in Section \ref{sec:Introduction}.
For effective comparisons, $x_i$ is drawn from $N(5,9)$ (see also \cite{ybarra2008small}) and $\psi_i \sim \text{Gamma}(4.5,2)$. Then, $\log Y_i= 3 x_i + v_i$, $\log y_i=\log Y_i+e_i$, and $W_i=x_i+u_i$. Here, $v_i \sim N(0,\sigma^2_v)$, $e_i \sim N(0,\psi_i)$, and $u_i \sim N(0,C_i)$. The sources of errors $v_i, e_i$, and $u_i$ are mutually independent. We let $\sigma^2_v=2$, and $C_i \in \{0, d\}$, where $d = 2$ or $4$ such that only $k\%$ of the $C_i$'s randomly receive $d$ and the rest receive $0$, where 
$k \in \{25, 50, 80, 100 \}$. The number of small areas are $m \in \{20,50,100\}$. 

This set-up of simulation has been previously used by \cite{ybarra2008small} which makes our results effectively comparable with theirs. We assume the total number of replications to be $R=2000$.       
We compare the performance of four predictors as follows, where we assume $Y_i$ is the truth: 
\begin{itemize}
\item[(1)] $y_i$: direct estimator, 
\item[(2)] $\tilde{\theta}_{i,\text{No-ME}}$: predictor without measurement error, 
\item[(3)] $\tilde{\theta}_{i,A}$: predictor A, and
\item[(4)] $\tilde{\theta}_{i,B}$: predictor B. 
\end{itemize}

For all the predictors, we substitute the estimated values of unknown parameters $\boldsymbol{\omega}$. 
The resulting predictors are listed in Table \ref{table.predictors}. Overall, the values of proposed predictor B are much closer to the truth $Y_i$ compared to the rest of other predictors.

We observe that when $C_i=0$ (no measurement error), the values of $\tilde{\theta}_{i,A}$, $\tilde{\theta}_{i,B}$, and $\tilde{\theta}_{i,\text{No-ME}}$ are identical. As the measurement error $C_i$ increases, the values of proposed predictor $\tilde{\theta}_{i,B}$ become much closer to the truth $Y_i$ rather than $\tilde{\theta}_{i,A}$. In the last row of Table \ref{table.predictors}, we report the average over the values of all small areas, which confirm the previous conclusion. 

\begin{table}[t!]
\centering
\caption{Comparison of predictors among all the small areas assuming $m=20$ and $k=50\%$. The numerical values are in the logarithmic scale.}
\label{table.predictors}
\renewcommand{\arraystretch}{1.5}
\vspace{0.25cm}
\begin{tabular}{@{} cccccc @{}}
\hline
$C_i$ & $Y_i$ & $y_i$ & $\tilde{\theta}_{i,\text{No-ME}}$ & 
$\tilde{\theta}_{i,A}$ & 
$\tilde{\theta}_{i,B}$  \\
\hline 
2 & 16.030 & 18.222 & 17.218 & 19.668 & 16.337  \\
0 &  34.128 & 38.389 & 36.726 & 36.726 & 36.726 \\
2 &   13.808 & 20.581 & 15.308 & 19.826 & 14.595 \\
2 &  -2.514  & 3.224  & 0.751 & 5.094 & -0.987 \\
0 &  9.524 & 14.283 & 11.011  & 11.011 & 11.011  \\
0 &  6.371 & 13.469 & 8.172 & 8.172  & 8.172  \\
2 &  10.209 & 14.919 & 11.995 & 16.771 & 10.640 \\
2 &  18.659 & 21.433 & 19.917 & 23.553 & 18.665 \\
0 &  14.746 & 18.158 & 15.858 & 15.858 & 15.858 \\
0 &  11.059 & 16.921 & 12.780 & 12.780 & 12.780 \\
2 &  26.719 & 32.409 & 29.597 & 34.329 & 27.178 \\
0 &  21.941 & 29.260 & 24.298 & 24.298 & 24.298 \\
2 &  1.420  & 8.516  & 4.125 & 9.875  & 4.291  \\
0 &  14.547 & 18.936 & 15.751 & 15.751 & 15.751 \\
2 &  13.990 & 18.599 & 16.078 & 20.317 & 14.602 \\
2 & 15.504 & 22.758 & 17.554 & 23.219 & 15.844  \\
2 &  16.491 & 20.650 & 18.181 & 22.292 & 16.827 \\
2 &  12.325 & 15.297 & 13.927 & 16.460 & 12.934 \\
0 &  16.549 & 20.563 & 17.501 & 17.501 & 17.501  \\
2 & 25.688 & 28.582 & 27.296 & 30.386 & 26.005  \\
Avg & 14.860 & 19.758 & 16.702 & 19.194 & 15.951  \\
\hline
\end{tabular}
\end{table}

We compare the performance of predictors based on the empirical MSE defined as follows: 
\begin{equation} \label{EMSE}
\text{EMSE}(\tilde{\theta}_i) = \frac{1}{R}\sum\limits_{r=1}^{R}\Big[\tilde{\theta}_{i}^{(r)}-Y_{i}^{(r)}\Big]^2,
\end{equation}
where $\tilde{\theta}_i$ is the predictor for $Y_i$, and $R$ is the total number of replications. Based on the results given in Table \ref{table.MSEs}, predictor B is superior to the rest of other predictors. Additionally, we make comparisons with the estimated MSE ($\hat{R}_{1i}$) and jackknife estimator ($mse_J$).
When $C_i=0$, the empirical MSE's for $\tilde{\theta}_{i,A}$, $\tilde{\theta}_{i,B}$, and $\tilde{\theta}_{i,\text{No-ME}}$ are identical, and when $C_i=2$, the empirical MSE's for $\tilde{\theta}_{i,B}$ are much smaller than the empirical MSE's for $\tilde{\theta}_{i,A}$. Overall, $\hat{R}_{1i}$ and $mse_J$ are much larger than the empirical MSE of predictor B. In the last row of the Table, we report the average over the values of all small areas. 

For further evaluation, we give the ratio of average MSE of predictor $\tilde{\theta}_i$ to average MSE of direct estimator $y_i$ in Table \ref{table.ratioMSEs}. When $C_i=0$, the ratio of average MSE's for $\tilde{\theta}_{i,A}$, $\tilde{\theta}_{i,B}$, and $\tilde{\theta}_{i,\text{No-ME}}$ are identical. When $C_i=2$, this ratio is much smaller for $\tilde{\theta}_{i,B}$ compared to $\tilde{\theta}_{i,A}$ and $\tilde{\theta}_{i,\text{No-ME}}$. Note that, since predictor A is substantially biased (see Section \ref{sec:Predictors}), its ratio of MSE to the direct one is very large.

\begin{table}[t!]
\centering
\caption{Comparison of empirical MSE of predictors among all the small areas assuming $m=20$ and $k=50\%$. The numerical values are in the logarithmic scale.}
\label{table.MSEs}
\renewcommand{\arraystretch}{1.5}
\vspace{0.25cm}
\begin{tabular}{@{} ccccccc @{}}
\hline
$C_i$ & EMSE($y_i$) & EMSE($\tilde{\theta}_{i,\text{No-ME}}$) & 
EMSE($\tilde{\theta}_{i,A}$) & EMSE($\tilde{\theta}_{i,B}$) & $\hat{R}_{1i}(\tilde{\theta}_{i,B})$ & $mse_J(\tilde{\theta}_{i,B})$ \\
\hline
2 &  39.620   &  37.307    &  42.255    &  35.659 & 81.028 & 83.227 \\
0 &  82.052  &   76.862    &   76.862   &   76.862 & 136.723 & 137.929 \\
2 & 48.272    &  35.470   &  45.134    &  34.809  & 84.564 & 67.638 \\
2 &  11.322    &  5.430   &  14.972    &   3.467 & 4.237 &  8.242 \\
0 & 34.702    &  25.503  &  25.503    &  25.503 & 39.084 & 38.530 \\
0 &  34.375   &  19.240  &   19.240   &   19.240 & 27.859 & 27.626 \\
2 & 35.133   &  27.569   &  39.131    &  25.354 & 56.307 & 52.168 \\
2 & 46.958   &  42.890  &  50.917    &  40.981 & 89.584 & 90.375 \\
0 &  41.084   &  34.251  &   34.251   &   34.251 & 57.158 & 57.389 \\
0 &  39.715   &  28.266   &   28.266   &   28.266 & 34.986 & 34.773 \\
2 & 70.133   &  64.302  &  72.810    &  59.415 & 121.389 & 117.451 \\
0 &  65.184  &  53.869   &   53.869   &   53.869 & 82.810 & 83.095 \\
2 &  23.345   &  11.770  &  26.675   &   15.912 & 22.204 & 23.868 \\
0 & 43.877   &  33.907  &  33.907    &  33.907 & 60.551 & 60.349 \\
2 & 42.366   &  36.899 &  44.901    &  35.275 & 71.884 & 70.593 \\
2 &  51.120   &  39.963  &   50.437   &   37.296 & 81.169 & 83.368 \\
2 &  45.980   & 40.585  &   48.605   &   37.410 & 82.652 & 85.104 \\
2 &  35.189   &  34.129 &   36.818   &   31.586 & 70.900 & 71.963 \\
0 & 47.121   &  37.576   &  37.576    &  37.576 & 70.871 & 71.125 \\
2 &  61.381   &  57.999   &   64.864   &  55.730 & 118.087 & 120.666 \\
Avg &  44.946   & 37.189  &    42.350  &   36.119 & 69.702  &  69.274 \\
\hline
\end{tabular}
\end{table}

\begin{table}[t!]
\centering
\caption{Ratio of average MSE of predictor $\tilde{\theta}_i$ (A, B, or No-ME) to average MSE of direct estimator $y_i$, where $m=20$.}
\label{table.ratioMSEs}
\renewcommand{\arraystretch}{1.5}
\vspace{0.25cm}
\begin{tabular}{@{} c|ccc @{}}
\hline
Error & \multicolumn{3}{c}{Ratio of MSEs} \\ 
\hline
$C_i \, (k=50\%)$ & $\tilde{\theta}_{i,\text{No-ME}}$ & $\tilde{\theta}_{i,A}$ & $\tilde{\theta}_{i,B}$ \\ 
 0 & 0.006 & 0.006 & 0.006 \\
 2 & 0.003  & 14.548 & 2.270e-05 \\
\hline
\end{tabular}
\end{table}

\begin{figure}[t!]
\centering
 \includegraphics[scale=0.45]{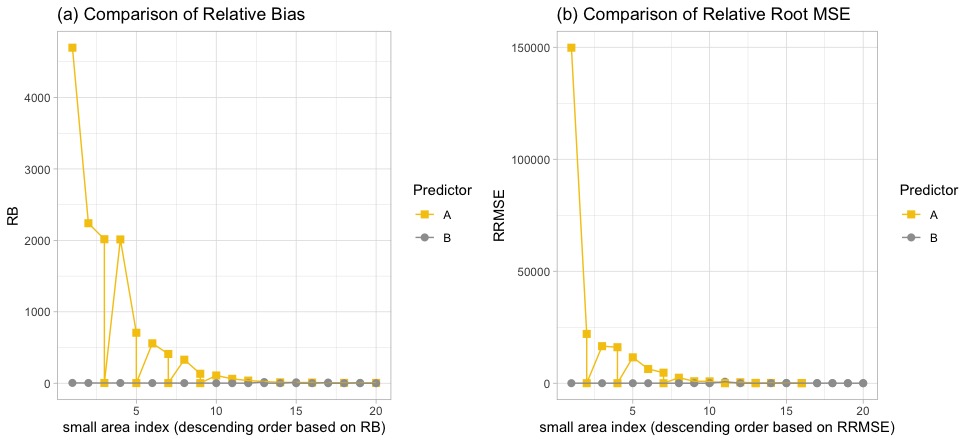} 
\caption{(a) Plot of RB for two predictors A and B. (b) Plot of RRMSE for two predictors A and B. We assume $m=20$ and $k=50\%$.}
\label{figure.RBRRMSE}
\end{figure}

We also provide the relative bias (RB) and the relative root mean squared error (RRMSE) of the main competitors; i.e. predictor A and predictor B in Figure \ref{figure.RBRRMSE}. 
These quantities can be defined as follows:
\begin{align*}
\text{RB} (\tilde{\theta}_i) =\frac{E[\tilde{\theta}_i-Y_i]}{Y_i}, \quad \text{and} \quad \text{RRMSE} (\tilde{\theta}_i) = \frac{\{E[\tilde{\theta}_i-Y_i]^2\}^{1/2}}{Y_i}.
\end{align*}   
Based on the results in Figure \ref{figure.RBRRMSE}, we observe that the RB's and RRMSE's of $\tilde{\theta}_{i,B}$ are very closely centered around 0, but this is not the case for the $\tilde{\theta}_{i,A}$. 

For the sake of completeness, we compare the predictors over all possible values of $k$ in Table S2.1 of the Supplementary Material for $m \in \{20,50,100\}$. The comparisons are based on the empirical MSE's which are averaged by values of $C_i$'s and confirm the previous arguments.

In order to compare the performance of our proposed prediction intervals for predictor B with the direct estimator, we compare the coverage probabilities and expected lengths of four prediction intervals as follows:

\begin{itemize}
\item[(1)] Direct: $[y_i - z_{1-\alpha/2} \, \sqrt{\psi_i}, \, y_i + z_{1-\alpha/2} \, \sqrt{\psi_i}]$,
\item[(2)] Estimated MSE: $[\tilde{\theta}_{i,B} - z_{1-\alpha/2} \, \sqrt{\hat{R}_{1i}(\tilde{\theta}_{i,B})}, \, \tilde{\theta}_{i,B} + z_{1-\alpha/2} \, \sqrt{\hat{R}_{1i}(\tilde{\theta}_{i,B})}]$,

\item[(3)] Jackknife: $[\tilde{\theta}_{i,B} - z_{1-\alpha/2}  \, \sqrt{mse_J(\tilde{\theta}_{i,B})}, \, \tilde{\theta}_{i,B} + z_{1-\alpha/2}  \, \sqrt{mse_J(\tilde{\theta}_{i,B})}]$, and
\item[(4)] Bootstrap: $[\ell_{i,\alpha}, \, u_{i,\alpha}]$ given in (\ref{eq.BootCI}).
\end{itemize}

\noindent We consider nominal coverage $100(1-\alpha)\%=90\%$, $95\%$, and $99\%$. The results are given in Table \ref{table.comparing_CP&L}. Overall, the bootstrap method outperforms  the other three methods.

The coverage probabilities for bootstrap method are close to the nominal coverage, thus the intervals more accurately include the truth. Prediction intervals based on the direct method, estimated MSE, and jackknife are not reliable as they do not approximately follow the normal theory. Additionally, the latter two usually suffer from coverage problem because of the choice of MSE estimator and small values of $m$.

\begin{table}[ht]
\centering
\caption{Comparison of coverage probabilities and mean of the log lengths for the prediction intervals across all the methods.
The results for the lengths are averaged over all the small areas. Additionally, we assume $k=50\%$ for all the cases.}
\label{table.comparing_CP&L}
\renewcommand{\arraystretch}{1.5}
\vspace{0.25cm}
\begin{tabular}{@{} cccccc @{}}
\hline
Nominal & Small Areas & Direct & Estimated MSE & Jackknife & Bootstrap  \\ 
Coverage & $m$ &  &  &  &  \\ 
\hline
$90\%$ & 20 & 0.520 & 0.450 & 0.420 & 0.990  \\
&& (15.383) & (12.195) & (13.646) & (30.463) \\
& 50 & 0.688 & 0.596 & 0.718 & 0.928 \\
&& (18.483) & (19.754) & (16.708) & (32.969) \\  \hline
$95\%$ & 20 & 0.560 & 0.450 & 0.420 & 1.000  \\
&& (15.561) & (12.374) & (13.824) & (37.314) \\
& 50 & 0.696 & 0.596 & 0.764 & 0.948 \\
&& (18.662) & (19.932) & (16.887) & (36.755) \\ \hline
$99\%$ & 20 & 0.560 & 0.500 & 0.500 & 1.000 \\
&&  (15.836) & (12.649) & (14.099) & (38.346) \\
& 50 & 0.736 & 0.596 & 0.791 & 0.992 \\
&& (18.936) & (20.207) & (17.161) & (37.692) \\
\hline
\end{tabular}
\end{table}

\section{Applications to Census Bureau's Data Sets} 
\label{sec:Realdata}

In this Section, we describe the steps used to apply the preceding theory to census data sets. 
The first application is related to the Census of Governments,  where we illustrate our methodology at the state level. 
The second application is related to the Small Area Income and Poverty Estimates (SAIPE) Program, where we illustrate our methodology at the county level. 

\subsection{Census of Governments} \label{subsec:CoG}

The purpose of the Census of Governments is providing periodic and comprehensive statistics about governments and governmental activities, and it covers all the states and local governments in the United States. 
Data are obtained on government organizations, finances, and employment and include location, type, and characteristics of local governments and officials; see \url{https://www.census.gov/econ/overview/go0100.html} for  further information. 

Since 1957 the United States Census Bureau collects information from governmental units for years ending in $2$ and $7$.  
Here, we utilize data from $2007$ and $2012$ with $49$ states of the Continental United States (excluding Hawaii and District of Columbia) as our small areas of interest.

We define the parameter of interest $\theta_i$ to be the mean number of full-time employees per government at state $i$ from the 2012 data set. We define the covariate to be the corresponding mean from the 2007 data set. To define the response $y_i$, we select sample of sizes $4000$ and $8000$ governmental units from the 2012 data set. Similarly, we construct the covariate $W_i$ from an independent sample of $40,000$ and $80,000$ governmental units selected from the 2007 data set. 

The distributions of response and covariate are displayed in Figures S3.1 and S3.2 of the Supplementary Material for sample sizes $4000$ and $8000$.
We observe skewed patterns in both the average number of full-time employees from 2007 and 2012 which motivates our proposed framework. Before a logarithmic transformation, both variables fail the normality assumption, and the normality assumption is more justified after the logarithmic transformation. 

The measurement error variances $C_i$'s for the covariates are obtained from a Taylor series approximation because of the logarithmic transformation, and the formula of variance in simple random sampling without replacement is used per each state in the original scale. Additionally, we used Taylor series approximation for the $\psi_i$.
We assume the sampling variances to be known throughout the estimation procedure.
Based on our proposed framework, we find the empirical predictor $\tilde{\theta}_{i,B}$. 

\begin{figure}[t!]
\centering 
 \includegraphics[width=0.75\textwidth]{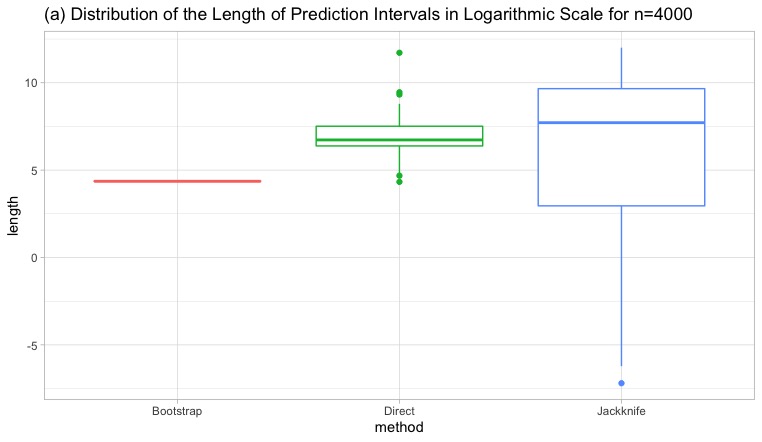} \\
 \includegraphics[width=0.75\textwidth]{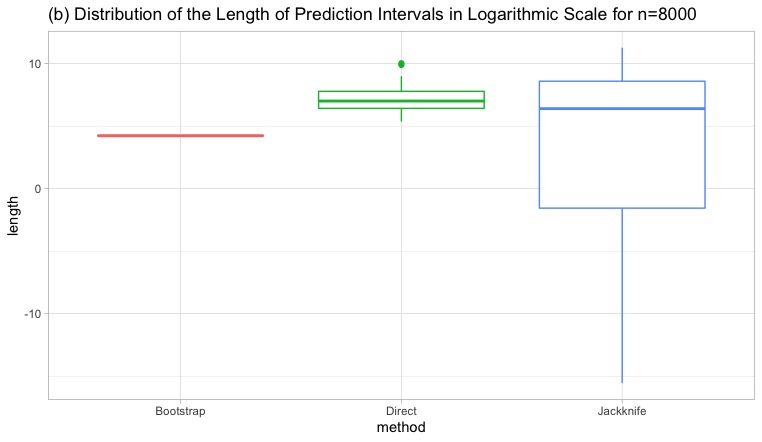} 
\caption{box-plots of prediction interval lengths with $1-\alpha=0.95$ based on three methods for the Census of Governments assuming (a) 4000 and (b) 8000  sample sizes.}
\label{figure.CI_CoG}
\end{figure}

We construct prediction intervals based on three methods of ``Direct", ``Jackknife", and ``Bootstrap". The box-plots of prediction interval lengths are given in Figure \ref{figure.CI_CoG}. We observe that the distribution of lengths based on non-parametric bootstrap method has less variation in comparison with both direct and jackknife methods. Additionally, the descriptive statistics of lengths for $95\%$ prediction intervals are given in Table \ref{table.des_stats_CoG}. We clearly observe more stable results for the descriptive statistics under the bootstrap method.

\begin{table}[t!]
\centering
\caption{Descriptive statistics for the log lengths of 95$\%$ prediction intervals from the Census of Governments. The results are computed using data from all the small areas.}
\label{table.des_stats_CoG}
\renewcommand{\arraystretch}{1.5}
\vspace{0.25cm}
\begin{tabular}{@{} cccccccc @{}}
\hline
Sample Size & Method & Minimum & $25\%$ & Median & Mean & $75\%$ & Maximum \\
\hline
4000 & Bootstrap & 4.320 &  4.350 &   4.360  &   4.370 &  4.380  &  4.410 \\
& Direct & 4.330  & 6.380  &  6.730   &  6.970  & 7.510  & 11.700 \\
& Jackknife & -7.200  & 2.950  & 7.710  &  5.840  &  9.660 &  12.000  \\
8000 & Bootstrap & 4.160  & 4.210 &  4.230  &  4.220  & 4.250  &  4.270 \\
& Direct &  5.370 &  6.420  & 7.000  &  7.150  & 7.780  & 10.000 \\
& Jackknife &  -15.600  & -1.560  & 6.400  &  3.760 &  8.590  & 11.300 \\
\hline
\end{tabular}
\end{table}

\subsection{Small Area Income and Poverty Estimates Program} \label{subsec:SAIPE}

The U.S. Census Bureau's SAIPE program uses \cite{fay1979estimates} model to  produce model-based estimates of income and poverty at the state and county levels for various age groups. Since 2005, they use the data from the American Community Survey (ACS) in the modeling. Prior to 2005, data from the Current Population Survey were used. The ACS is the largest U.S. household survey and it almost covers 3.5 million addresses per year. 

Despite the large sample size of the ACS, the 1-year direct estimates of the number of related school-aged (5-17 year old) children in poverty are highly variable for many small counties. Thus, SAIPE program uses the \cite{fay1979estimates} model to borrow strength from covariates such as log number of food-stamp participants, log number of IRS child exemptions in households in poverty, log number of related children aged 5-17 in poverty from previous census, etc. to improve the direct estimate of the logarithm of the single-year ACS poverty counts as the dependent variable.   
For more information on the SAIPE program, see the SAIPE web page at \url{https://www.census.gov/programs-surveys/saipe.html}.

Recent research by \cite{huang2012empirical} and  \cite{franco2015borrowing} suggests that the most recent previous 5-year ACS estimates instead of outdated census results can be used as a covariate (see, \cite{arima2017}). However, 5-year ACS estimates are subject to the sampling errors rather than the census results. Thus, it is appropriate to use \cite{fay1979estimates} where covariates are measured with errors and a log transformation is required. As a consequence, our proposed predictor is well-suited for this scenario. 

Here, we assume the parameter of interest is the total population for whom poverty status is determined for the age of 5 to 17 years in 2018 at the county level. The covariate is the corresponding total of 5-year aggregated values; i.e. 2013--2017 at the county level. Additionally, we use the 2018  SNAP data set as a separate covariate which is measured without error. 

We summarize these variables as follows:
\begin{itemize}
\item $y_i$: 1-year  ACS (2018) estimates for $i=1,...,827$.
\item $W_i$: Aggregated 5-year  ACS (2013--2017) estimates, measured with errors, and
\item $x_i$: 2018 SNAP data set measured without error as a separate covariate. 
\end{itemize}

Unlike the 5-year aggregated ACS, the information for counties with population less than 65,000 are not publicly released for a single-year ACS. We successfully linked the same m=827 counties across the resources, and we only provide predictions for these publicly available counties. 
In Figure S3.3 (a) given in the Supplementary Material, we display the scatter plots of these counties. 

We observe that  both 5-year ACS and SNAP are highly correlated with the response variable $y$. The skewness in the original scales diminishes after a logarithmic transformation (see Figure S3.3 (b) in the Supplementary Material).  
To obtain the variance of estimates, we use the 90$\%$ margin of error given in the data set for both the covariate and the response variable. Afterwards, we use a Taylor series approximation to obtain the variance of the logarithm of estimates. 

In this Section, we intend to study three small area models as follows:
\begin{itemize}
\item[(1)] Measurement error model predictor: For this model, we assume the response variable is the 1-year ACS and the covariate is the 5-year ACS (measured with error). Because of skewed patterns in the data set, we need to use a logarithmic transformation. For this purpose, we use predictor B.
\item[(2)] Fay-Herriot model predictor: For this model, we assume the response variable is the 1-year ACS and the covariate is the SNAP information (measured without error). Because of skewed patterns in the data set, we again use a logarithmic transformation. For this purpose, we use predictor $\tilde{\theta}_{i,B}^{0}$, the so-called FHeblup afterwards.    
\item[(3)] Direct Estimator: The results are solely based on the single-year ACS estimates.
\end{itemize}

\begin{figure}[t!]
\centering
 \includegraphics[width=0.85\textwidth]{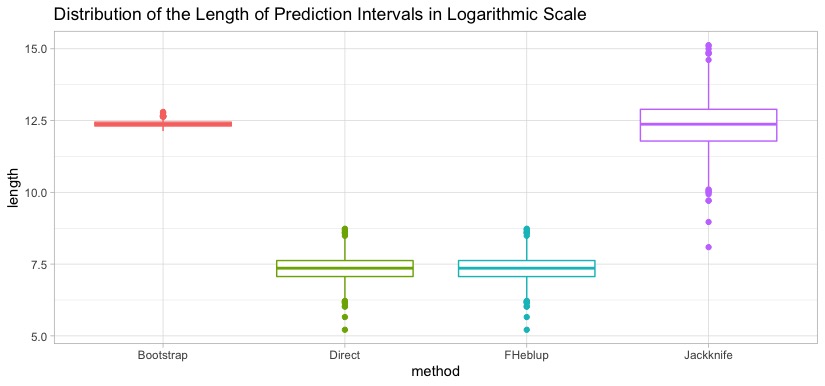} 
\caption{Box-plots of prediction interval lengths with $1-\alpha=0.95$ for the SAIPE data set. Note that the results are in the logarithmic scale.}
\label{figure.CI_SAIPE}
\end{figure}

In Figure \ref{figure.CI_SAIPE}, we compare the distribution of prediction interval lengths for all the counties based on non-parametric bootstrap, jackknife, direct, and FHeblup. We observe that the box-plot for the prediction interval lengths based on the non-parametric bootstrap method has a stable distribution. 

Additionally, box-plots of direct and FHeblup are similar due to the close values of direct estimators and FHeblups (see Figure S3.4 in the Supplementary Material). This means the values of $\tilde{\theta}_{i,B}^{0}$ are close to $z_i$, which requires $\psi_i \approx 0$. This can be confirmed by the range of $\psi_i$, which is (6.579\,e-07, 6.843\,e-03). The values of mean, median, and standard deviation of the lengths for all the prediction intervals  are given in Table \ref{table.des_stats_SAIPE}.   

\begin{table}[ht]
\centering
\caption{Mean, median, and standard deviation for the log lengths of 95$\%$ prediction intervals from the SAIPE data set. The results are computed using data from all the small areas.}
\label{table.des_stats_SAIPE}
\renewcommand{\arraystretch}{1.5}
\vspace{0.25cm}
\begin{tabular}{@{} cccc @{}}
\hline
Method & Median & Mean & Standard Deviation \\ [0.15cm]
\hline
Bootstrap & 12.360  &   12.400  & 0.112  \\
Direct & 7.358  &   7.342  &  0.453 \\
FHeblup & 7.357  &  7.342  & 0.453 \\
Jackknife & 12.370  &   12.340 &   0.905  \\
\hline
\end{tabular}
\end{table}

\section{Discussion and Future Work} \label{sec:discussion}

We propose a new predictor for the skewed response variable under Fay-Herriot model when the covariate is measured with error. Our set-up can be easily extended to the multivariate covariate case, and some of the steps in this regard are given in the Supplementary Material. 
While the proposed method is for area level model, it would be possible to consider an extension to unit level models, which is left to a future work.

This modeling framework can be of interest for government agencies and survey practitioners dealing with right skewed response variables and covariates that are measured with errors. Our proposed predictor is unbiased and can perform uniformly better than the direct estimator and the alternative predictor gien in the literature by \cite{mosaferi2020+}. Further, we derive an approximation of the MSE of predictor, which does not perform well for constructing prediction intervals in terms of coverage. Thus, we develop nonparametric bootstrap prediction intervals. 

Prediction intervals based on nonparametric bootstrap techniques are easy and allow a good coverage property. In particular, they can be more suitable for real applications as we use the available data sets for generating resamples. It might be of interest to investigate other resampling methods such as parametric bootstrap methods or extend the earlier works of log-MSPE by \cite{jiang2016unified} to construct prediction intervals and make comparisons among them for our modeling framework. 

	\section*{Supplementary Material}
 The online Supplementary Material includes technical details, proofs of the theorems, and additional numerical results. 
 
 	\section*{Acknowledgements}
	
	We would like to thank an associate editor and anonymous reviewers who made excellent comments and suggestions that helped us to improve the paper. 

\bibliographystyle{ims}
\bibliography{Bibliography}


\newpage
\clearpage

\setcounter{secnumdepth}{0}

\begin{center}
SUPPLEMENTARY MATERIAL FOR ``AN UNBIASED PREDICTOR FOR SKEWED RESPONSE VARIABLE WITH MEASUREMENT ERROR IN COVARIATE"

\vspace{0.25cm}
Sepideh Mosaferi, Malay Ghosh, and Shonosuke Sugasawa
\vspace{0.25cm}

{\it University of Massachusetts Amherst, University of Florida} \\
    {\it and Keio University}
\end{center}

\vspace{1cm}

This supplementary material is structured as follows. Regularity conditions are given in Section S1. Further simulation and application results are given in Sections S2 and S3, respectively. Section S4 contains multivariate covariate setup.    
We provide proofs of theorems and some technical derivations in Sections S5 and S6. 

\vspace{1cm}

\section{S1 \quad Regularity conditions} \label{sec.conditions}

We give three regularity conditions as follows:

\vspace{0.25cm}
\noindent\textbf{Condition 1.} $\{(y_i,W_i,\psi_i)\}$ is a sequence of independent and identically distributed random vectors, and there exist positive constants $\psi_L$ and $\psi_U$ such that $0< \psi_L \leq \inf_{1 \leq i \leq m} \psi_i \leq \sup_{1 \leq i \leq m} \psi_i \leq \psi_U < \infty$ for $i = 1,...,m$.

\vspace{0.25cm}
\noindent\textbf{Condition 2.} $\boldsymbol{\omega}=(\beta_0,\beta_1,\sigma^2_v)' \in \Theta$ where $\Theta$ is a  compact set such that $\Theta \subset (\mathbb{R},\mathbb{R},\mathbb{R}_{+})$ and $\hat{\boldsymbol{\omega}} \xrightarrow{P} \boldsymbol{\omega}$.

\vspace{0.25cm}
\noindent\textbf{Condition 3.} (i) $\tilde{U}(\boldsymbol{\omega})$ exists almost surely in probability and $E\{\tilde{U}(\boldsymbol{\omega})\}=0$. (ii) $\tilde{U}'(\boldsymbol{\omega})$ is a continuous function where $E\{\tilde{U}'(\boldsymbol{\omega})\}$ is uniformly bounded away from zero. (iii) $E\{|\tilde{U}(\boldsymbol{\omega})|^{4+\delta} \}$, $E\{|\tilde{U}'(\boldsymbol{\omega})|^{4+\delta} \}$, and $E\{\sup_{c \in (-\epsilon,\epsilon)}|\tilde{U}''(\boldsymbol{\omega})|^{4+\delta} \}$ are uniformly bounded under some $\epsilon>0$ and $\delta>0$.

\section{S2 \quad Further simulation results} \label{sec.simulation}

\setcounter{equation}{0}
\setcounter{table}{0}
\setcounter{figure}{0}
\def\theequation{S2.\arabic{equation}}
\def\thesection{S\arabic{section}}
\def\thetable{S2.\arabic{table}}
\def\thefigure{S2.\arabic{figure}}  

In this Section, we provide the empirical MSE of predictors as well as $\hat{R}_{1i}$ and $mse_J$ for multiple values of small areas related to the simulation Section of the paper. The results are listed in Table \ref{table.MSEsallk's}.

\begin{table}[t!]
\centering
\caption{Empirical MSE as well as $\hat{R}_{1i}$ and $mse_J$ of predictors averaged by the values of $C_i$ for all possible values of $k$. We assume $m \in \{20,50,100\}$, and the numerical values are in the logarithmic scale.}
\label{table.MSEsallk's}
\renewcommand{\arraystretch}{1.5}
\vspace{0.25cm}
\begin{tabular}{@{} cccccccc @{}}
\hline
$k$ & $C_i$ & EMSE($y_i$) & EMSE($\tilde{\theta}_{i,\text{No-ME}}$) & 
 EMSE($\tilde{\theta}_{i,A}$) & EMSE($\tilde{\theta}_{i,B}$) & $\hat{R}_{1i}(\tilde{\theta}_{i,B})$ & $mse_J(\tilde{\theta}_{i,B})$ \\
 \hline
\multicolumn{8}{c}{$m=20$} \\
25 & 0 & 44.678 & 33.385 & 33.385 & 33.385 & 54.275 & 57.497 \\
   & 2 &  48.742 & 37.031 & 50.733 & 38.925 & 90.202 & 80.191 \\
50 & 0 & 48.514 & 38.684 & 38.684 & 38.684 & 63.755 & 63.852 \\
   & 2 & 42.568 & 36.193 & 44.793 & 34.408 & 73.667 & 72.889  \\
80 & 0 & 47.321 & 40.189 & 40.189 & 40.189 & 62.135 & 65.010 \\
   & 2 &  43.966 & 37.164 & 46.019 & 35.167 & 79.058 & 78.690 \\
100 & 2 & 45.600 & 38.923 & 47.445 & 37.089 & 81.210 & 78.617 \\ \hline
\multicolumn{8}{c}{$m=50$} \\
25 & 0 & 48.779 & 37.851 & 37.851 & 37.851 & 66.433 & 68.337 \\
   & 2 &  47.535 & 37.699 & 49.199 & 39.195 & 90.517 & 89.474 \\
50 & 0 & 47.239 & 37.227 & 37.227 & 37.227 & 63.285 & 65.029 \\
   & 2 &  50.841 & 41.380 & 52.488 & 42.141 & 96.313 & 94.237 \\
80 & 0 & 49.722 & 40.308 & 40.308 & 40.308 & 71.310 & 74.125 \\
   & 2 & 48.903 & 41.514 & 50.419 & 40.469 & 90.451 & 88.179 \\
100 & 2 & 48.613 & 42.374 & 49.999 & 41.082 & 89.500 & 88.080 \\ \hline
\multicolumn{8}{c}{$m=100$} \\
25 & 0 & 40.311 & 28.807 & 28.807 & 28.807 & 49.891 & 52.212 \\
   & 2 & 49.897 & 38.359 & 51.599 & 40.349 & 92.106 & 89.288 \\
50 & 0 & 46.091 & 35.481 & 35.481 & 35.481 & 62.508 & 64.791  \\
   & 2 & 39.885 & 29.954 & 41.941 & 31.010 & 71.395 & 68.307 \\   
80 & 0 & 55.042 & 46.172 & 46.172 & 46.172 & 77.347 & 77.868 \\
   & 2 &  39.988 & 33.310 & 41.700 & 32.434 & 72.332 & 71.142 \\
100 & 2  &  42.977 & 36.106 & 44.535 & 35.132 & 78.235 & 77.868 \\  \hline
\end{tabular}
\end{table}

\section{S3 \quad Further application results} \label{sec.application}

\setcounter{equation}{0}
\setcounter{table}{0}
\setcounter{figure}{0}
\def\theequation{S3.\arabic{equation}}
\def\thesection{S\arabic{section}}
\def\thetable{S3.\arabic{table}}
\def\thefigure{S3.\arabic{figure}}  

\begin{figure}[t!]
\centering
 \includegraphics[width=1\textwidth]{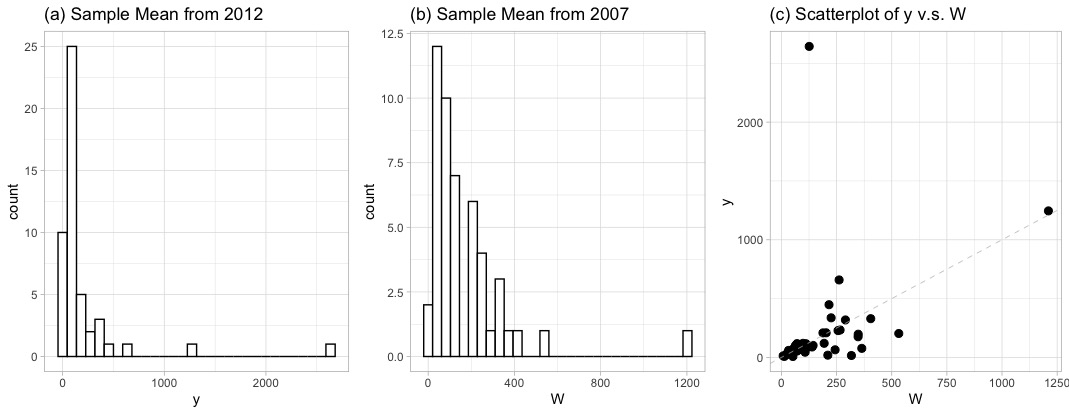} \\
 \includegraphics[width=1\textwidth]{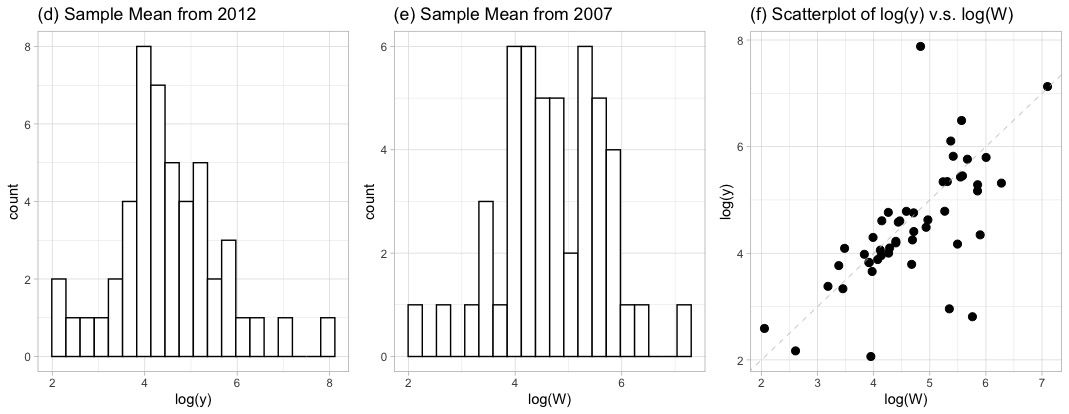} 
\caption{Histograms for the Census of Governments based on 4000 sample size. In both plots (a) and (b), the distributions of covariate and response are highly skewed to the right side. After transformations and in plots (d) and (e), we observe a stabilized distribution. Plots (c) and (f) display the regression relationship between the response variable and covariate before and after transformation.}
\label{figure.CoGtrans_4000}
\end{figure}

\begin{figure}[t!]
\centering 
 \includegraphics[width=1\textwidth]{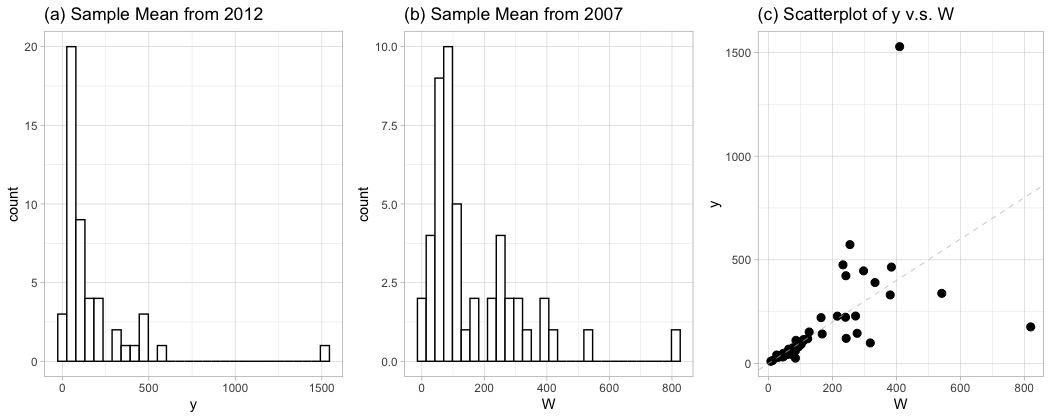} \\
 \includegraphics[width=1\textwidth]{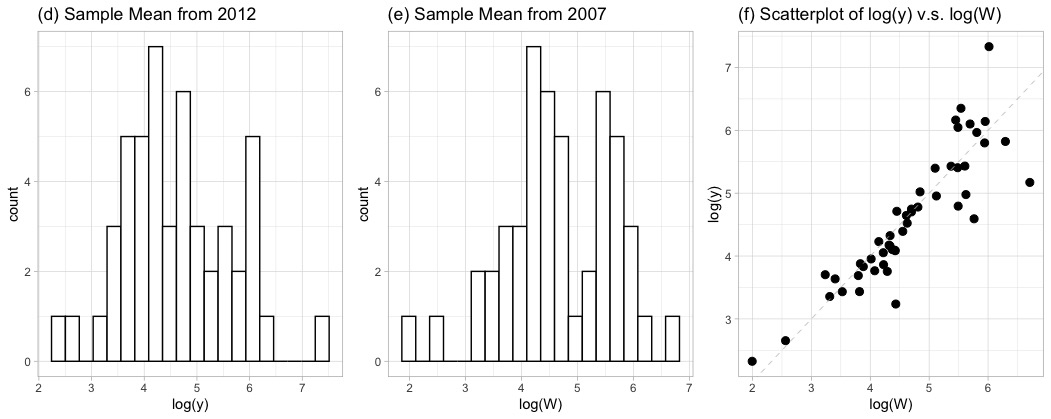} 
\caption{Histograms for the Census of Governments based on 8000 sample size. In both plots (a) and (b), the distributions of covariate and response are highly skewed to the right side. After transformations and in plots (d) and (e), we observe a stabilized distribution. Plots (c) and (f) display the regression relationship between the response variable and covariate before and after transformation.}
\label{figure.CoGtrans_8000}
\end{figure}

\begin{figure}[t!]
\centering 
 \includegraphics[width=1.1\textwidth]{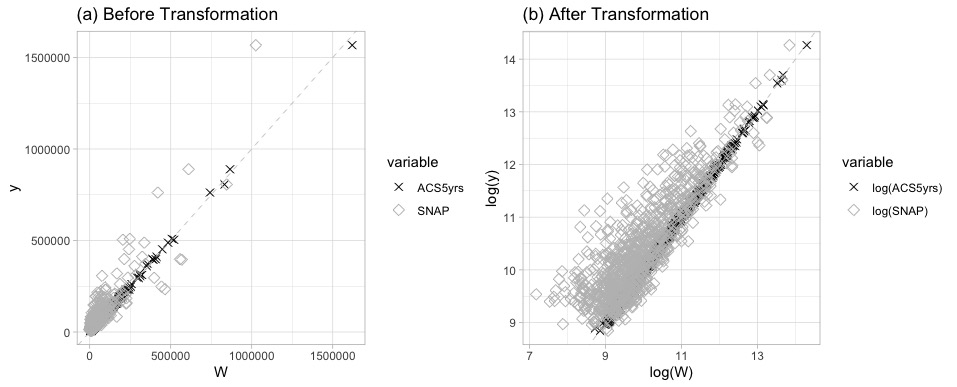}
\caption{Scatter plots of response variable versus covariates (a) before and (b) after transformations for the SAIPE data set.}
\label{figure.scatter_SAIPE}
\end{figure}

\begin{figure}[t!]
\centering 
 \includegraphics[width=0.8\textwidth]{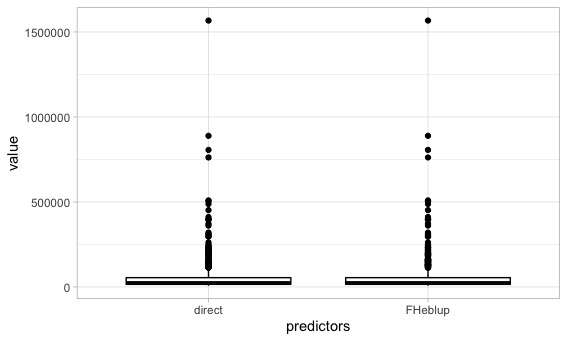}
\caption{Box-plots of direct and FHeblup predictors for the SAIPE data set.}
\label{figure.covs_SAIPE}
\end{figure}

In this Section, we provide three figures related to the application Section of the paper.
Figures \ref{figure.CoGtrans_4000} and \ref{figure.CoGtrans_8000} depict the distributions of the Census of Governments based on 4000 and 8000 sample sizes. Figure \ref{figure.scatter_SAIPE} shows the scatter plots for the SAIPE data set. Figure \ref{figure.covs_SAIPE} displays the box-plots of two predictors from the SAIPE data set.

\section{S4 \quad Multivariate extension} \label{sec.multiple_covs}

In this Section, we give some details of formulation and forms of predictors A and B for multivariate covariate set-up. Let's assume the following hierarchical set-up
\begin{align*}
z_i| \phi_i & \sim  N(\phi_i,\psi_i) \\
\phi_i & \sim N(\beta_0+ \boldsymbol{\beta}' \boldsymbol{x}_i , \sigma^2_v) \\
\boldsymbol{W_i} & \sim \textit{MVN}(\boldsymbol{x}_i, \boldsymbol{C}_i),
\end{align*}
where $\boldsymbol{\beta}'=(\beta_1,\beta_2,...,\beta_p)$, $\boldsymbol{x}_i=(x_{i1},x_{i2},...,x_{ip})'$, $\boldsymbol{W}_i=(W_{i1},W_{i2},...,W_{ip})'$, and $\boldsymbol{C}_i=\text{diag}(C_{i1},...,C_{ip})$.

The parameter of interest is $\theta_i=\exp(\beta_0+\boldsymbol{\beta}' \boldsymbol{x}_i+v_i)$. Following the same derivations given in the manuscript, predictor A can be defined as
\begin{equation*}
\tilde{\theta}_{i,A} = \exp(\tilde{\gamma}_i z_i + (1-\tilde{\gamma}_i)(\beta_0+\boldsymbol{\beta}' \boldsymbol{W}_i)+\tilde{\gamma}_i \psi_i/2),  
\end{equation*}
where $\tilde{\gamma}_i=(\boldsymbol{\beta}' \boldsymbol{C}_i \boldsymbol{\beta}+\sigma^2_v)/S_i(\boldsymbol{\beta},\sigma^2_v)$ and $S_i(\boldsymbol{\beta},\sigma^2_v)=(\boldsymbol{\beta}' \boldsymbol{C}_i \boldsymbol{\beta}+\sigma^2_v+\psi_i)$. Predictor B can be defined as
$\tilde{\theta}_{i,B}= \tilde{\theta}_{i,A} \exp(-\frac{1}{2} d_i)$, where $d_i=2 \psi_i \boldsymbol{\beta}' \boldsymbol{C}_i \boldsymbol{\beta}/S_i(\boldsymbol{\beta},\sigma^2_v)$. The vector of unknown parameters can be estimated along the same lines of the manuscript.

\section{S5 \quad Proofs of theorems} \label{sec.proofs}
\setcounter{equation}{0}
\setcounter{table}{0}
\setcounter{figure}{0}
\def\theequation{S5.\arabic{equation}}

In this Section, we provide proofs of theorems.

\vspace{0.25cm}
\noindent \textbf{Proof of Theorem 1:} 
\begin{flalign} \label{E_predA}
E[\tilde{\theta}_{i,A}] & = E[\exp(\tilde{\gamma}_i z_i)] E[\exp\{ (1-\tilde{\gamma}_i) (\beta_0+\beta_1 W_i) \}] \exp(\tilde{\gamma}_i \psi_i/2) && \nonumber\\
& = \exp [\tilde{\gamma}_i (\beta_0+\beta_1 x_i+ \frac{1}{2} \tilde{\gamma}_i^2 (\sigma^2_v + \psi_i)) ] && \nonumber \\
& \quad \times \exp[(1-\tilde{\gamma}_i) (\beta_0+\beta_1 x_i + \frac{1}{2} (1-\tilde{\gamma}_i)^2 (\beta_1^2 C_i))] \exp(\tilde{\gamma}_i \psi_i/2) && \nonumber\\
& = \exp [\beta_0+\beta_1 x_i+\frac{1}{2} \{\tilde{\gamma}_i^2 (\sigma^2_v+\psi_i)+ (1-\tilde{\gamma}_i)^2(\beta_1^2 C_i)+\tilde{\gamma}_i \psi_i\}].
\end{flalign}
Next, we simplify
\begin{flalign} \label{secondterm}
& \tilde{\gamma}_i^2 (\sigma^2_v+\psi_i)+ (1-\tilde{\gamma}_i)^2 \beta_1^2 C_i+\tilde{\gamma}_i \psi_i && \nonumber\\
& \qquad = \tilde{\gamma}_i^2 (\sigma^2_v+\psi_i+\beta_1^2 C_i) -2 \tilde{\gamma}_i \beta_1^2 C_i + \beta_1^2 C_i + \tilde{\gamma}_i \psi_i && \nonumber\\
& \qquad = S_i^{-1}(\beta_1,\sigma^2_v) (\beta_1^2 C_i+\sigma^2_v)^2 - 2 \tilde{\gamma}_i \beta_1^2 C_i + \beta_1^2 C_i + \tilde{\gamma}_i \psi_i && \nonumber\\
& \qquad = \tilde{\gamma}_i (\beta_1^2 C_i+\sigma^2_v) -2 \tilde{\gamma}_i \beta_1^2 C_i + \beta_1^2 C_i + \tilde{\gamma}_i \psi_i && \nonumber\\
& \qquad = \tilde{\gamma}_i (\sigma^2_v+\psi_i)+(1-\tilde{\gamma}_i) \beta_1^2 C_i.
\end{flalign}
The result follows from (\ref{E_predA}) and (\ref{secondterm}).
\vspace{0.25cm}

\noindent\textbf{Proof of Theorem 2:}
\noindent We use the equations from the expressions (\ref{eq:finalunbiasdscore}) of the manuscript to find the matrix $I_{\boldsymbol{\omega}}$ as follows

\begin{equation*}
I_{\boldsymbol{\omega}} = \begin{bmatrix}
var(\tilde{U}_1(\boldsymbol{\omega})) & cov(\tilde{U}_1(\boldsymbol{\omega}), \tilde{U}_2(\boldsymbol{\omega})) & cov(\tilde{U}_1(\boldsymbol{\omega}),\tilde{U}_3(\boldsymbol{\omega})) \\
cov(\tilde{U}_1(\boldsymbol{\omega}), \tilde{U}_2(\boldsymbol{\omega})) & var(\tilde{U}_2(\boldsymbol{\omega})) & cov(\tilde{U}_2(\boldsymbol{\omega}), \tilde{U}_3(\boldsymbol{\omega})) \\
cov(\tilde{U}_1(\boldsymbol{\omega}),\tilde{U}_3(\boldsymbol{\omega})) & cov(\tilde{U}_2(\boldsymbol{\omega}),\tilde{U}_3(\boldsymbol{\omega})) & 
var(\tilde{U}_3(\boldsymbol{\omega}))
\end{bmatrix}.
\end{equation*}
The elements of the matrix are as follows
\begin{flalign*}
 var(\tilde{U}_1(\boldsymbol{\omega})) & = \sum_{i=1}^{m} S_i^{-1}(\beta_1,\sigma^2_v), && \\
 cov(\tilde{U}_1(\boldsymbol{\omega}), \tilde{U}_2(\boldsymbol{\omega})) & = \sum_{i=1}^{m} S_i^{-2}(\beta_1,\sigma^2_v) cov[W_i \tau_i(\beta_0,\beta_1), \tau_i(\beta_0,\beta_1)] && \\
& = \sum_{i=1}^{m} S_i^{-2}(\beta_1,\sigma^2_v) E[(W_i-x_i+x_i)\tau_i^2(\beta_0,\beta_1)] = \sum_{i=1}^{m} S_i^{-1}(\beta_1,\sigma^2_v) x_i, && \\
 cov(\tilde{U}_1(\boldsymbol{\omega}),\tilde{U}_3(\boldsymbol{\omega})) & = 0, && \\
 var(\tilde{U}_2(\boldsymbol{\omega})) & = \sum_{i=1}^{m} S_i^{-2}(\beta_1,\sigma^2_v) var[W_i \tau_i(\beta_0,\beta_1)] + \beta_1^2 \sum_{i=1}^{m} S_i^{-4}(\beta_1,\sigma^2_v) C_i^2 \,  var[\tau_i^2(\beta_0,\beta_1)] &&\\
& \quad + 2 \beta_1 \sum_{i=1}^{m} S_i^{-3}(\beta_1,\sigma^2_v) C_i cov[W_i \tau_i(\beta_0,\beta_1), \tau_i^2(\beta_0,\beta_1)].
\end{flalign*}
\noindent Note that we have 
\begin{itemize}
\item[(i)] $var[W_i \tau_i(\beta_0,\beta_1)]= (x_i^2+C_i) S_i(\beta_1,\sigma^2_v)+ \beta_1^2 C_i^2$,
\item[(ii)] $var[\tau_i^2(\beta_0,\beta_1)]= 2 S_i^{2}(\beta_1,\sigma^2_v)$, and
\item[(iii)] $cov[W_i \tau_i(\beta_0,\beta_1), \tau_i^2(\beta_0,\beta_1)]= - 2 \beta_1 C_i S_i(\beta_1,\sigma^2_v)$.
\end{itemize} 
Therefore, 
\begin{flalign*}
var(\tilde{U}_2(\boldsymbol{\omega})) & = \sum_{i=1}^{m} S_i^{-2}(\beta_1,\sigma^2_v) (\sigma^2_v + \psi_i) C_i + \sum_{i=1}^{m} S_i^{-1}(\beta_1,\sigma^2_v) x_i^2 = \sum_{i=1}^{m} S_i^{-1}(\beta_1,\sigma^2_v) (x_i^2 + \tilde{\sigma}^2_{ci}), &&\\
 cov(\tilde{U}_2(\boldsymbol{\omega}), \tilde{U}_3(\boldsymbol{\omega})) & = \frac{1}{2} cov[\sum_{i=1}^{m} S_i^{-1}(\beta_1,\sigma^2_v) W_i \tau_i(\beta_0,\beta_1) + \beta_1 \sum_{i=1}^{m} S_i^{-2}(\beta_1,\sigma^2_v) C_i \tau_i^2(\beta_0,\beta_1), &&\\
& \sum_{i=1}^{m} S_i^{-1}(\beta_1,\sigma^2_v) \tau_i^2(\beta_0,\beta_1)] &&\\
& = \frac{1}{2} \sum_{i=1}^{m} S_i^{-3}(\beta_1,\sigma^2_v) cov[W_i \tau_i(\beta_0,\beta_1), \tau_i^2(\beta_0,\beta_1)] &&\\
& \quad + \frac{1}{2} \beta_1 \sum_{i=1}^{m} S_i^{-4}(\beta_1,\sigma^2_v) C_i var[\tau_i^2(\beta_0,\beta_1)] = - \beta_1 \sum_{i=1}^{m} S_i^{-2}(\beta_1,\sigma^2_v) C_i &&\\
& \quad + \beta_1 \sum_{i=1}^{m} S_i^{-2}(\beta_1,\sigma^2_v) C_i=0,  &&\\
 var(\tilde{U}_3(\boldsymbol{\omega})) & = \frac{1}{4} \sum_{i=1}^{m} S_i^{-4}(\beta_1,\sigma^2_v) var[\tau_i^2(\beta_0,\beta_1)] = \frac{1}{2} \sum_{i=1}^{m} S_i^{-2}(\beta_1,\sigma^2_v).
\end{flalign*}
\noindent As a final result, we get 
\begin{equation*}
I_{\boldsymbol{\omega}} = \begin{bmatrix}
\sum_{i=1}^{m}S_i^{-1}(\beta_1,\sigma^2_v) & \sum_{i=1}^{m} S_i^{-1}(\beta_1,\sigma^2_v) x_i & 0 \\
\sum_{i=1}^{m} S_i^{-1}(\beta_1,\sigma^2_v) x_i  & \sum_{i=1}^{m} S_i^{-1}(\beta_1,\sigma^2_v) (x_i^2+ \tilde{\sigma}^2_{ci} )& 0\\
0 & 0 & \frac{1}{2} \sum_{i=1}^{m} S_i^{-2}(\beta_1,\sigma^2_v)
\end{bmatrix}.
\end{equation*}

\section{S6 \quad Details of derivations for $\hat{R}_{1i}$} \label{sec.R1i}
\setcounter{equation}{0}
\setcounter{table}{0}
\setcounter{figure}{0}
\def\theequation{S6.\arabic{equation}}
\noindent Recall that 
$R_{1i} := M_{1i}(\boldsymbol{\omega}) M_{2i}(\boldsymbol{\omega})$. In order to estimate $R_{1i}$, one can define
\begin{align} \label{expansion}
& E[M_{1i}(\hat{\boldsymbol{\omega}}) M_{2i}(\hat{\boldsymbol{\omega}}) - M_{1i}(\boldsymbol{\omega}) M_{2i}(\boldsymbol{\omega})]^2  := E [\{ M_{1i}(\hat{\boldsymbol{\omega}}) - M_{1i}(\boldsymbol{\omega}) \} 
\{ M_{2i}(\hat{\boldsymbol{\omega}}) - M_{2i}(\boldsymbol{\omega}) \} \nonumber \\
& \qquad + M_{2i}(\boldsymbol{\omega}) (M_{1i}(\hat{\boldsymbol{\omega}}) - M_{1i}(\boldsymbol{\omega})) + M_{1i}(\boldsymbol{\omega}) (M_{2i}(\hat{\boldsymbol{\omega}}) - M_{2i}(\boldsymbol{\omega}))]^2 \nonumber \\
& \quad = E[\{M_{1i}(\hat{\boldsymbol{\omega}})- M_{1i}(\boldsymbol{\omega}) \}^2 
\{M_{2i}(\hat{\boldsymbol{\omega}})- M_{2i}(\boldsymbol{\omega}) \}^2] \nonumber \\
& \qquad + M_{2i}^2(\boldsymbol{\omega}) E[M_{1i}(\hat{\boldsymbol{\omega}})- M_{1i}(\boldsymbol{\omega})]^2
+ M_{1i}^2(\boldsymbol{\omega}) E[M_{2i}(\hat{\boldsymbol{\omega}})- M_{2i}(\boldsymbol{\omega})]^2 \nonumber \\
& \qquad + 2 E[(M_{1i}(\hat{\boldsymbol{\omega}})- M_{1i}(\boldsymbol{\omega}))^2 (M_{2i}(\hat{\boldsymbol{\omega}})- M_{2i}(\boldsymbol{\omega}))] M_{2i}(\boldsymbol{\omega}) \nonumber \\
& \qquad +2 E[(M_{1i}(\hat{\boldsymbol{\omega}})- M_{1i}(\boldsymbol{\omega})) (M_{2i}(\hat{\boldsymbol{\omega}})- M_{2i}(\boldsymbol{\omega}))^2] M_{1i}(\boldsymbol{\omega}) \nonumber \\
& \qquad +2 M_{1i}(\boldsymbol{\omega}) M_{2i}(\boldsymbol{\omega}) E[(M_{1i}(\hat{\boldsymbol{\omega}})- M_{1i}(\boldsymbol{\omega})) (M_{2i}(\hat{\boldsymbol{\omega}})- M_{2i}(\boldsymbol{\omega}))].
\end{align}

\noindent Application of the Cauchy-Schwarz inequality yields
\begin{align*} 
\text{(i)} \quad & E[\{ M_{1i}(\hat{\boldsymbol{\omega}})- M_{1i}(\boldsymbol{\omega}) \}^2 \{ M_{2i}(\hat{\boldsymbol{\omega}})- M_{2i}(\boldsymbol{\omega}) \}^2] \\
& \quad \leq E^{1/2} [M_{1i}(\hat{\boldsymbol{\omega}})- M_{1i}(\boldsymbol{\omega})]^4 E^{1/2} [M_{2i}(\hat{\boldsymbol{\omega}})- M_{2i}(\boldsymbol{\omega})]^4 
 = O(1) O(m^{-1}) = O(m^{-1}), \\
\text{(ii)} \quad & E[\{ M_{1i}(\hat{\boldsymbol{\omega}})- M_{1i}(\boldsymbol{\omega}) \}^2 \{ M_{2i}(\hat{\boldsymbol{\omega}})- M_{2i}(\boldsymbol{\omega}) \}] \\
& \quad \leq E^{1/2} [M_{1i}(\hat{\boldsymbol{\omega}})- M_{1i}(\boldsymbol{\omega})]^4 E^{1/2} [M_{2i}(\hat{\boldsymbol{\omega}})- M_{2i}(\boldsymbol{\omega})]^2 = O(1) O(m^{-1/2}) = O(m^{-1/2}), \\
\text{(iii)} \quad & E[\{ M_{1i}(\hat{\boldsymbol{\omega}})- M_{1i}(\boldsymbol{\omega}) \} \{ M_{2i}(\hat{\boldsymbol{\omega}})- M_{2i}(\boldsymbol{\omega}) \}^2] \\
& \quad \leq E^{1/2} [M_{1i}(\hat{\boldsymbol{\omega}})- M_{1i}(\boldsymbol{\omega})]^2 E^{1/2} [M_{2i}(\hat{\boldsymbol{\omega}})- M_{2i}(\boldsymbol{\omega})]^4 = O(1) O(m^{-1}) = O(m^{-1}), \\
\text{(iv)} \quad & E[M_{2i}(\hat{\boldsymbol{\omega}})- M_{2i}(\boldsymbol{\omega})]^2 = O(m^{-1}), \quad \text{and} 
\end{align*}

\begin{align*} 
\text{(v)} \quad & E[\{ M_{1i}(\hat{\boldsymbol{\omega}})- M_{1i}(\boldsymbol{\omega}) \} \{ M_{2i}(\hat{\boldsymbol{\omega}})- M_{2i}(\boldsymbol{\omega}) \}] \\
& \quad \leq E^{1/2} [M_{1i}(\hat{\boldsymbol{\omega}})- M_{1i}(\boldsymbol{\omega})]^2 E^{1/2} [M_{2i}(\hat{\boldsymbol{\omega}})- M_{2i}(\boldsymbol{\omega})]^2 = O(1) O(m^{-1/2}) = O(m^{-1/2}). 
\end{align*}
\noindent Thus, we conclude that only the term $M^2_{2i}(\boldsymbol{\omega}) E[M_{1i}(\hat{\boldsymbol{\omega}})- M_{1i}(\boldsymbol{\omega})]^2$ from expression (\ref{expansion}) needs to be estimated. Therefore, the estimator of $R_{1i}$ is the expression of  $\hat{R}_{1i}$ given in the manuscript.

\end{document}